\documentclass{aa}  

\usepackage{graphicx,lscape,bm}
\usepackage{txfonts}
\newcommand{\1}{{\,\sc{i}}}
\newcommand{\2}{{\,\sc{ii}}}
\newcommand{\3}{{\,\sc{iii}}}
\defcitealias{Polimera2022a}{P22}
\usepackage{orcidlink}
\hypersetup{hidelinks}
\begin{document}

   \title{Recovering the properties of the interstellar medium through integrated spectroscopy: Application to the $z\sim0$ ECO volume-limited star-forming galaxy sample}
   \titlerunning{Recovering ISM properties in $z\sim0$ star-forming galaxies through integrated spectroscopy}

   \author{V.\ Lebouteiller\inst{1}\orcidlink{0000-0002-7716-6223}
          \and
          C.\ T. Richardson\inst{2}\orcidlink{0000-0002-3703-0719}
           \and
           M.\ S. Polimera\inst{3,4}\orcidlink{0000-0001-6162-3963}  
           \and
           D.\ S. Carr\inst{4}\orcidlink{0009-0007-5354-3817}
           \and
           Z.\ L. Hutchens\inst{4,5}\orcidlink{0000-0002-8574-5495}
          \and
           S.\ J. Kannappan\inst{4}\orcidlink{0000-0002-3378-6551}
           \and
            L.\ Ramambason\inst{6}\orcidlink{0000-0002-9190-9986}
           \and
           A.\ J. Moffett\inst{7}\orcidlink{}
           \and 
           M.\ Varese\inst{1}\orcidlink{0009-0004-6699-8341}
           \and
           S.\ C. Madden\inst{1}\orcidlink{0000-0003-3229-2899}
          }

   \institute{$^1$: Université Paris-Saclay, Université Paris Cité, CEA, CNRS, AIM, 91191, Gif-sur-Yvette, France\\
              \email{vianney.lebouteiller@cnrs.fr}\\
              $^2$: Elon University, 100 Campus Drive, Elon, NC 27278, USA \\
              $^3$: Center for Astrophysics | Harvard \& Smithsonian, 60 Garden Street, Cambridge, MA 02138, USA \\
              $^4$: Department of Physics \& Astronomy, University of North Carolina at Chapel Hill, Chapel Hill, NC 27599, USA \\
              $^5$: Department of Physics \& Astronomy, University of North Carolina Asheville, Asheville, NC 28804, USA \\
              $^6$: Institut fur Theoretische Astrophysik, Zentrum für Astronomie, Universität Heidelberg, Albert-Ueberle-Str. 2, 69120 Heidelberg,
Germany \\
                $^7$: Department of Physics and Astronomy, University of North Georgia, 3820 Mundy Mill Road, Oakwood GA 30566, USA\\
             }

   \date{Received Sep. 3rd, 2024; accepted Jan. 24, 2025}

 
  \abstract
  {Deriving physical parameters from integrated galaxy spectra is paramount to interpret the cosmic evolution of the star formation, chemical enrichment, and energetic processes at play. Previous studies have highlighted the power of  interstellar medium tracers but also the associated complexities that can be captured only through sophisticated modeling approaches. }
  {We developed modeling techniques to characterize the ionized gas properties in the subset of $2052$ star-forming galaxies from the volume-limited, dwarf-dominated, $z\sim0$ ECO catalog (stellar mass range $M_*\sim10^{8-11}$\,M$_\odot$). Our study sheds light on the internal distribution and average values of parameters such as the metallicity, ionization parameter, and electron density within galaxies. }
   {We used the MULTIGRIS statistical framework to evaluate the performance of various models using strong lines as constraints. The reference model involves physical parameters distributed as power laws with free parameter boundaries. Specifically, we used combinations of 1D photoionization models (i.e., considering the propagation of radiation toward a single cloud) to match optical H\2\ region lines, in order to provide probability density functions of the inferred parameters. }
   {The inference predicts nonuniform physical conditions within galaxies. The integrated spectra of most galaxies are dominated by relatively low-excitation gas with a metallicity around $0.3$\,Z$_\odot$. Using the average metallicity in galaxies, we provide a new fit to the mass-metallicity relationship which is in line with direct abundance method determinations from the low-metallicity calibrated range up to high-metallicity stacks.  The average metallicity shows a weakly bimodal distribution which may be due to external (e.g., refueling of non-cluster early-type galaxies) or internal processes (higher star-formation efficiency in metal-rich regions). The specific line set used for inference affects the results and we identify potential issues with the use of the [S\2] line doublet.
  }  
   {Complex modeling approaches may capture diverse physical conditions within galaxies but require robust statistical frameworks. Such approaches are limited by the inherent 1D model database as well as caveats regarding the gas geometry. Our results highlight, however, the possibility to extract useful and significant information from integrated spectra.
   }


   \maketitle
%

   \section{Introduction}

Spectroscopic diagnostics of the interstellar medium (ISM) in galaxies hold tremendous diagnostic power, for example, on the star-formation rate (SFR), gas masses, or the fraction of ionizing radiation due to active galactic nuclei (AGNs). Yet we do not know precisely how to interpret spatially unresolved spectra by accounting for and modeling the complex mechanisms that produce the  observed, integrated, emission.
As we accumulate more high-$z$ observations 
as well as wide-field and all-sky spectroscopy lacking spatial resolution,
it becomes urgent to design a modeling framework to derive robust and reliable physical parameter distributions highlighting galaxy evolution mechanisms. 
Furthermore, integrated observations may encompass enough information to actually recover such distributions, thereby enabling some of the power of integral-field spectroscopy (IFS) without actually performing it. 

\begin{figure*}
\centering
   \includegraphics[width=18cm,height=10cm]{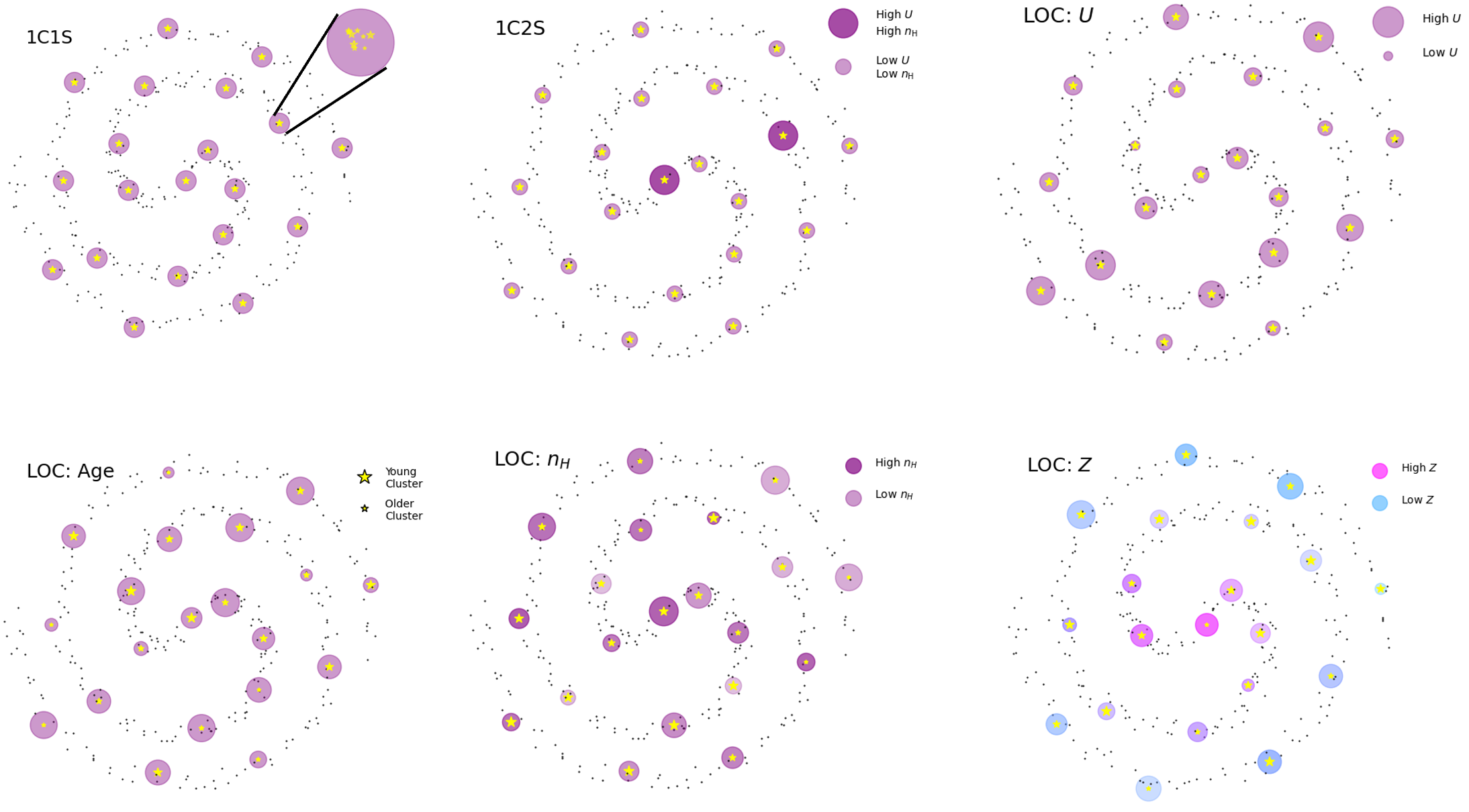}
   \caption{Illustration of topologies using multicomponent models (1C1S and 1C2S) and integrated distributions (LOC). 1C1S assumes a stellar population described with a single age associated with an ISM component with uniform conditions. 1C2S uses the same stellar population hypothesis as 1C1S but enables two distinct sets of uniform ISM conditions. LOC models gradually consider parameters as power-law distributions. }
   \label{fig:topos}
\end{figure*}

Historically, long-slit spectroscopy or integrated spectroscopy have been used to probe some kinds of average physical conditions in galaxies or to identify dominant excitation sources. Integrated line ratios are, for instance, commonly interpreted using 1D photoionization grids (i.e., assuming a single cloud illuminated by a unique radiation source -- or possibly several co-spatial radiation sources -- with a plane-parallel or spherical geometry) in order to trace the gas electron density and pressure, the ionization parameter ($U$), the metallicity ($Z$), SFR, or excitation mechanism diagnostics (see, e.g., \citealt{Kewley2019a}). Such a hypothesis may reflect either a single ``representative'' H\2\ region or, equivalently, an ensemble of H\2\ regions with similar properties (see illustration in Fig.\,\ref{fig:topos}). 

It is generally difficult to evaluate the reliability and actual meaning of the quantities derived thusly and it is therefore crucial to understand potential biases and selection effects due to the hypothesis of a single emission component versus\ the combined emission of various components and physical mechanisms (see, e.g., \citealt{Sorba2015a,Sorba2018a} for stellar mass determinations). 
Star-forming galaxies are indeed known to include the following: \begin{itemize}
\item gas that follows a density distribution related to turbulence, self-gravitation, and rotational support (e.g., \citealt{Khullar2021a}); 
\item metallicity variations within galaxies in the form of gradients but also higher order variations (e.g., \citealt{Poetrodjojo2018a,Williams2022a,Nakajima2024a});
\item molecular clouds, some associated with recent star formation (e.g., \citealt{Tacconi2020a,Saintonge2022a}); 
\item a collection of H\2\ regions following some luminosity function (e.g., \citealt{Santoro2022a}), with some of them leaking ionizing photons (e.g., \citealt{DellaBruna2021b});
\item stellar age gradients (e.g., \citealt{Riggs2024a}); and
\item additional excitation sources such as Wolf-Rayet stars, high-mass X-ray binaries, or AGNs, with the stellar clusters and AGN that may or not be co-spatial, resulting in coincident or non-coincident geometries \citep{Richardson2022a}.
\end{itemize}
  
In some rare cases (e.g., guided by imagery or spatially resolved spectroscopy), it is possible to describe a specific star-forming region or even a specific galaxy as a single dominant stellar cluster surrounded by ISM shells, which enables the use of full 3D or pseudo 3D models (with photoionization and chemistry) with arbitrary geometries (e.g., M$^3$; \citealt{Jin2022a}, PyCloudy; \citealt{Morisset2013a,Fitzgerald2020a}). Three-dimensional Monte-Carlo radiative transfer codes are also useful in that they can handle potentially complex geometries and density structures (e.g., \citealt{Baes2003a,Baes2011a,deLooze2014a}) with promising avenues toward a fully self-consistent 3D model \citep{Romero2023a}. Despite these possible improvements, the geometry is never a free parameter in the models that involve chemistry and photoionization. 
It must also be added that radiation magneto-hydrodynamical simulations are now able to solve large chemical networks (e.g., \citealt{Katz2022b,Katz2022c}), while simulation databases are increasingly available (e.g., \citealt{Katz2023a}) but the comparison to observations remains difficult due to the generally restricted parameter space (e.g., regarding cosmic ray ionization rates or the dust-to-gas mass ratio). 

Due to the difficulty in designing 3D models, single 1D models are frequently used, often in conjunction with modern statistical frameworks (e.g., \texttt{BEAGLE}; \citealt{Chevallard2016a}, \texttt{CIGALE}; \citealt{Burgarella2005a,Boquien2019a,Yang2022a}). It should be noted that 1D models can be somewhat optimized by considering  different geometries (e.g., spherical geometry with thin shells or filled spheres; e.g., \citealt{Stasinska2015a}), by including incomplete shells (matter- and density-bounded regions; e.g., \citealt{Pequignot2008a,Cormier2015a,Ramambason2020a,Ramambason2022a}), or by accounting for the flux-averaged integrated emission of evolving H\2\ regions (e.g., \citealt{Dopita2006a,Groves2008a,Pellegrini2019a}).

Another avenue consists in combining 1D models, either representing a few ``sectors'' around stellar clusters (e.g., \citealt{Pequignot2008a,Cormier2015a,Lebouteiller2017a,Cormier2019a,Madden2020a,Ramambason2022a}) or statistical distributions of many emitting components within a galaxy (e.g., \citealt{Richardson2014a,Richardson2016a,Richardson2019a,Lebouteiller2022a,Ramambason2024a,Marconi2024a,Varese2025a,Morisset2025a}). Such combinations provide useful alternatives to 3D models as long as projection effects are not an important issue.

Classically, the approach to model an integrated galaxy is often driven by the availability of tracers. In general, one prefers the simplest possible model (i.e., the smallest possible set of free parameters) that matches available observations, while more complex configurations (such as combinations of 1D models) are usually introduced out of necessity. This raises an important phenomenological question as to whether models should consider the following:
  \begin{itemize}
  \item An optimal model ``architecture'' (i.e., choice of physical and geometrical parameters) adapted to the data and preventing too much overfitting. The drawback is that different physical descriptions of the galaxy are solely considered based on what tracers are available, resulting in ``representative'' models that are often difficult to interpret.
  \item A model architecture driven by a -- possibly complex -- physical description of the galaxy, with the most likely parameter values inferred from available observations. This may result in weakly constrained model parameters when few tracers are available but the model itself remains identical with additional tracers.
\end{itemize}
\noindent In the present study, we start from the principle that the model architecture represents a physical object and is expected to be as robust as possible against the set of available tracers. 
For this we relied on combinations of 1D models (``topological models'' as pioneered in \citealt{Pequignot2002a,Pequignot2008a}), as they enable a high enough level of complexity that may approach the actual distribution of source and ISM clouds -- despite several biases and caveats -- while also being easily parameterized. In other words, we consider that the improvements enabled by such combinations compared to single 1D models largely compensate for potential biases. Apart from the physically motivated necessity to include distributions of components and parameters in order to extract specific parameters of interest related to galaxy evolution, we are also interested in actually recovering the intrinsic variation of physical conditions (metallicity, density, etc.), keeping in mind that IFS samples of dwarf galaxies are particularly small and that such indirect methods may provide promising alternatives.

Assuming such a modeling approach, it is essential to construct a reliable framework to compare models and observations. Probabilistic methods are most adapted as they remain useful when the set of tracers changes and/or when parameters are correlated. Full precomputed grids (including all potential parameter combinations) enable brute-force methods with a Bayesian likelihood calculated for every model and they are relatively quick to process large observation sets (e.g., \citealt{Blanc2015a,Thomas2016a}). However, for this work, we relied on on-the-fly Bayesian likelihood calculations within the statistical framework MULTIGRIS\footnote{\url{https://gitlab.com/multigris/mgris}} \citep{Lebouteiller2022a}. MULTIGRIS enables one to account for nuisance variables and is better adapted to a large number of parameters. For completeness, it must be mentioned that neural networks are also increasingly used to match model predictions and observations, especially in the case of a large number of parameters (e.g., \citealt{Kang2022a,Morisset2025a}). Furthermore, new deep learning methods enable model outcomes to be emulated in even faster ways than regular interpolation methods (e.g.,\,\citealt{Palud2023a}).

We focus on a star-forming galaxy sample extracted from the volume-limited Environmental COntext (ECO) survey \citep{Moffett2015a,Polimera2022a,Hutchens2023a}, which is complete into the dwarf galaxy regime and for which there should be minimal AGN contamination. Most of the galaxies are dwarf galaxies, but we can study statistically meaningful correlations between parameters and internal variations over a large mass range. Assuming that past versions of massive galaxies resemble low-mass galaxies in the present sample at $z\sim0$, we may interpret our results as a potential probe of evolutionary pathways. Specifically, we wish to examine the following: 1) the range of physical conditions present in the ECO sample under the most realistic hypothesis of multiple emitting components and how these conditions evolve as a function of statistically averaged quantities (e.g., as a function of metallicity); and 2) the link between physical parameters such as $U$ and $Z$, and their connection to galaxy evolution parameters such as SFR.

In Section\,\ref{sec:framework} we present the framework to assess the various model architectures used to model galaxies. We then present the application of this framework to the ECO star-forming galaxy sample with constraints from optical line spectroscopy (Sect.\,\ref{sec:application}). Results are described in Section\,\ref{sec:results} where we examine the following: 1) the influence of the line set used, and find potential issues with the [S\2] lines; 2) the differences between various model architectures, and find that architectures using statistical distributions outperform single 1D models; and 3) the main model fit parameters. Section\,\ref{sec:discussion} explores the following: 1) correlations between physical parameters, with strong correlations of all parameters with metallicity $Z$; and 2) the recovery of internal distributions within galaxies with an emphasis on the metallicity dispersion. We also discuss the potential existence and implication of a metallicity bimodality as well as the mass-metallicity relationship. For the latter, we do find evidence of a significant increase in metallicity between stellar masses $10^{9.5}-10^{10}$\,M$_\odot$.

\section{Modeling framework}\label{sec:framework}

\begin{figure*}\label{fig:decision}
\centering
   \includegraphics[width=18cm]{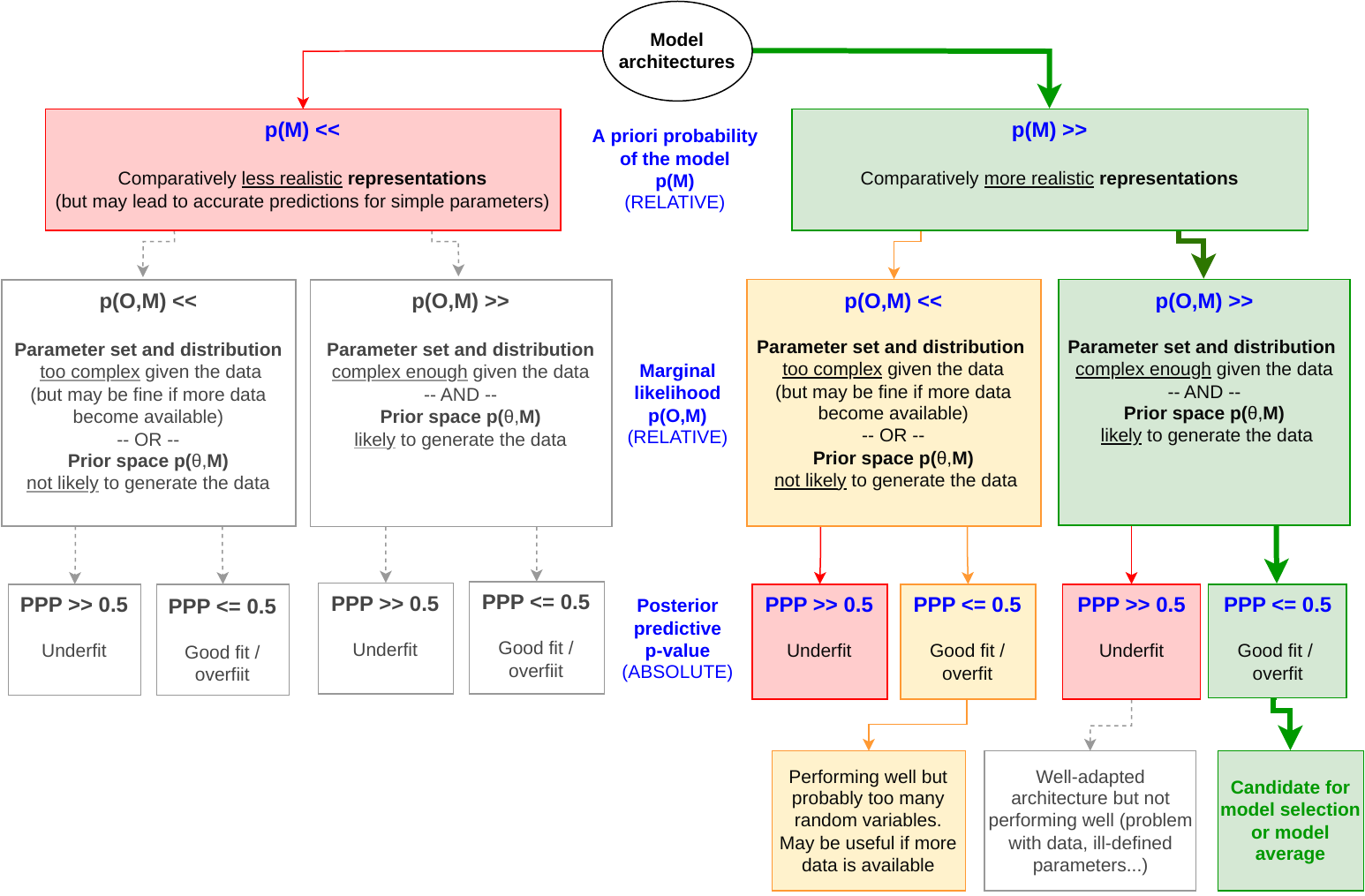}
   \caption{Decision tree for model architectures. The successive metrics are indicated in the middle. Green arrows and boxes indicate the path of maximum likelihood for model consideration. }
\end{figure*}

\subsection{Definition of model architectures (topological models)}

Photoionization models are used to describe parameters related to the sources (spectral energy distribution, luminosity), the ISM (density $n$, chemical composition, distance from source, etc.), and the gas excitation conditions (ionization parameter $U$). 
The simplest approach to model a full galaxy considers a single ``virtual'' stellar cluster representing all clusters in the galaxy with spherically symmetric ISM conditions. This single 1D model may be either interpreted as representing 1) the full galaxy with all excitation sources being co-spatial, resulting in ``effective'' (or ``representative'') galaxy-wide parameters, or 2) a collection of strictly identical 1D components (clusters surrounded with ISM) whose total luminosity amounts to that of the galaxy. Both interpretations are equivalent as long as the radiation transfer is controlled by the absorption of UV ionizing photons by the gas (classical H\2\ regions hypothesis described by the Str\"omgrem sphere assumption; \citealt{Osterbrock2006a}). We show in Figure\,\ref{fig:topos} (top left) an illustration of such models to mimic a galaxy's emitting components.

We may then consider a linear combination of two or few ``components,'' with each component representing a single 1D photoionization model. The combination may describe either 1) several ISM components surrounding a single stellar cluster (i.e., well adapted to single H\2\ regions with relatively dense and diffuse ``sectors'' around the young massive cluster; see, e.g., \citealt{Cormier2019a,Ramambason2022a}), 2) several stellar clusters each surrounded with identical ISM conditions (i.e., well adapted to the case of young SF regions and old stellar populations), or 3) any combination of the above. A critical caveat is that the linear combination assumes that the components are independent. Consequently, radiation escaping from one component does not affect other components. We will generally refer to these relatively simple architectures of one or a few 1D models as multicomponent models or ``$x$C$y$S,'' for $x$ clusters associated with $y$ ISM shell components around each cluster. 

As the number of components increases, it becomes necessary to tie them through a statistical distribution described by specific hyperparameters. This is both to keep a manageable number of free parameters as well as to consider a physically meaningful distribution. This is the motivation behind the locally optimally emitted cloud (LOC; \citealt{Ferguson1997a,Richardson2014a}) hypothesis that the observed emission is the result of strong selection effects due to the fact that some emission lines are brighter under certain conditions. Therefore, we may consider a large number of clouds whose model parameters (e.g., density, stellar cluster age) are distributed as power or normal laws defined and integrated between either fixed or free boundaries. We will generally refer to these more complex model architectures as ``integrated'' distributions or simply ``LOC.'' Progressively more complex LOC models are illustrated in Figure\,\ref{fig:topos}.

The LOC equation for an extensive observable $L$ (e.g., a line flux) for a series of parameter sets ${\bf p}=(p_0 ... p_n)$ gives
\begin{equation} L_{\rm tot} = \sum_{\bf p_{min}}^{\bf p_{max}} \prod_{i=0}^n \Phi({p_i}) I({\bf p}) \Delta({p_i}),
\end{equation} with $I({\bf p})$ the observable for a given parameter set, $\Delta({p_i})$ the grid parameter interval, and $\Phi(p_i)$ the weight associated with the parameter $p_i$ according to some statistical distribution (e.g., power law). For instance, the weight for a power-law distribution of some parameter $p$ would be \begin{equation}\label{eq:alpha}
\Phi(p) = \left\{
    \begin{array}{ll}
        10^{\alpha_p} & \mbox{if } p \in [p_{\rm min}, p_{\rm max}] \\
        0 & \mbox{}
    \end{array}
  \right\}.
  \end{equation}
  \noindent The weight for a parameter that does not follow any particular distribution and that is described instead by a single value $p_{\rm val}$  would be defined as \begin{equation}\label{eq:delta}
\Phi(p) = \delta_p = \left\{
    \begin{array}{ll}
        1 & \mbox{if } p = p_{\rm val} \\
        0 & \mbox{}
    \end{array}
  \right\}.
  \end{equation}
The combined weight for a parameter set ${\bf p}$ is then 
\begin{equation}
\Phi({\bf p}) = \prod_{i=0}^n \Phi(p_i).
\end{equation}
The free parameters are either the model parameters themselves ($p_{\rm val}$ for a single valued distribution) or the hyperparameters (e.g., $\alpha_{p}$, $p_{\rm min}$, $p_{\rm max}$ for a power-law distribution). 

For any parameter distribution considered, the average parameter value is a useful quantity to calculate, e.g., to compare to single 1D model approaches. The average parameter value is defined as \begin{equation}\label{eq:avg}
p_{\rm avg} = \frac{\sum_{\bf p_{min}}^{\bf p_{max}} \Phi(p) p}{\sum_{\bf p_{min}}^{\bf p_{max}} \Phi(p)},
\end{equation}
where $p$ is in log scale\footnote{Since the integration is performed in the logarithmic space for all parameters, it is natural to calculate the LOC average as the average of the logarithmic quantities. For instance, the combination of two models with all parameters being the same except for different densities $10^3$ and $10^0$\,cm$^{-3}$ compared to two models of densities $10^3$ and $10^{-1}$\,cm$^{-3}$ should lead to significantly different results which would be better reflected by the average of the log densities rather than the linear ones.}.

We emphasize that the distribution of components that we may recover using this method does not correspond to the distribution of what actually composes a galaxy but it is instead the luminosity-weighted distribution of the components that contribute to the optical emission line fluxes. For benchmark purposes, we compared the LOC hyperparameters inferred on-the-fly to results obtained with precomputed LOC grids, (i.e., with tabulated hyperparameters).  The inference of hyperparameters quickly becomes more efficient compared to pre-tabulated grids when the number of parameters increases and also provides a flexible framework with priors and potential nuisance variables.

\subsection{Statistical framework with \texttt{MULTIGRIS}}

While LOC distributions may be adapted to particular types of galaxies or regions within galaxies, there is often no prior knowledge as to what distributions should be considered. In the following, we propose one potential method to compare various architectures. We use the statistical framework provided by \texttt{MULTIGRIS} \citep{Lebouteiller2022a} which performs on-the-fly inference of combination of 1D models through Markov Chain Monte Carlo (MCMC) sampling. In the following, we refer to the model $\mathcal{M}$ as a model ``architecture'' defined by a certain distribution of parameters. The posterior probability distribution for a given model $\mathcal{M}$ is
defined as \begin{equation}
  p({\bm \theta}|\bm{O},\mathcal{M}) = \frac{p(\bm{O}|{\bm \theta},\mathcal{M}) p({\bm \theta}|\mathcal{M})}{ p(\bm{O}|\mathcal{M})},
\end{equation}
with $\bm{O}$ the data, ${\bm \theta}$ the parameters, $p(\bm{O}|{\bm \theta},\mathcal{M})$ the likelihood, $p({\bm \theta}|\mathcal{M})$ the prior probability, and $p(\bm{O}|\mathcal{M})$ the marginal likelihood.

Several LOC distributions have been implemented in \texttt{MULTIGRIS} (power laws, broken power laws, and normal distributions), with the ability to provide priors on the hyperparameters (slope, mean, standard deviation, etc.). Since most ISM models are usually too long to run for individual MCMC draws, the inference relies on the 1D model grid sampling, together with an interpolation method which can be either nearest neighbors or multidimensional linear interpolation. In practice, hyperparameters for LOC models are either continuous (slope $\alpha_p$ for the power law, mean, and standard deviation for a normal distribution) or using linear or nearest neighbor interpolation (boundaries $p_{\rm min,max}$). 
The first application of LOC distributions with \texttt{MULTIGRIS} was presented in \cite{Ramambason2024a} where it was required to explain the emission of the CO(1-0) emission in metal-poor galaxies, with the cloud depth in particular described by a broken power-law distribution. 

\subsection{Comparative and performance metrics}\label{sec:metrics}

Several metrics are important to consider when evaluating a model. In particular, we are interested in how well the model captures the data (``goodness'' of fit), which is well described by the posterior predictive $p$-value ($PPP$). The $PPP$ performs statistical tests of many simulated datasets from the model, using the parameters inferred from the observed data. The $PPP$ is then the proportion of these simulated test statistics that are more extreme than the test statistic calculated from the real data, and is defined as
\begin{equation}
p(\bm{O}_{\rm rep}|\bm{O})= \int_{\bm \theta} p(\bm{O}_{\rm rep}|{\bm \theta}) p({\bm \theta}|\bm{O}) d{\bm \theta},
\end{equation}
with $\bm{O}_{\rm rep}$ as generated sets of replicated observables (see, e.g., \citealt{Galliano2021a}). Ideally the $PPP$ should be around $0.5$, while values near $1$ imply a poor fit (underfit) and values near $0$ imply a an overfit.

Another important quantity is the marginal likelihood which integrates all parameter combinations from the prior space and therefore enables hypothesis testing (i.e., what is the simplest model adapted to the data), defined as: \begin{equation}
p(\bm{O}|\mathcal{M}) = \int_{\bm \theta}
p(\bm{O}|{\bm \theta},\mathcal{M}) p({\bm \theta}|\mathcal{M}) d{\bm \theta}.
\end{equation}
When comparing two models to each other, one may use the Bayes factor which is the ratio of the marginal likelihoods. However, it is necessary to consider the a priori probability of the models themselves, $p(\mathcal{M})$, describing how well the set of parameters is adapted to the object we wish to model (independently on the exact set of observations). The Bayes Factor for two models $\mathcal{M}_1$ and $\mathcal{M}_2$ then becomes
  \begin{equation}\label{eq:bf}
BF = \frac{ p(\bm{O}|\mathcal{M}_1) }{ p(\bm{O}|\mathcal{M}_2) } \frac{ p({\mathcal{M}_1}) }{ p({\mathcal{M}_2}) }.
\end{equation}

While the $PPP$ are easily calculated and interpreted in absolute ways, the marginal likelihood and the a priori probability of the models are much more complicated. The marginal likelihood is often too difficult to evaluate for simple random walkers as it is necessary to sample well enough the entire prior space. \texttt{MULTIGRIS} uses the Sequential Monte-Carlo method which runs a large number of small Markov Chains across the prior space in a series of steps until convergence to the posterior distribution \citep{DelMoral2012a}. This makes it possible to estimate the marginal likelihood for each model\footnote{While the Sequential Monte-Carlo method is particularly adapted to multi-model posterior distributions, we keep in mind that minor modes may be dropped during the resampling phase, effectively cropping the posterior distribution. Our inference runs use the largest possible number of particles to alleviate this issue.}. The a priori model probability $p(\mathcal{M})$ is difficult, if not impossible, to evaluate. By default one may simply consider that a set of models are equally plausible to represent a galaxy and therefore ignore these probabilities, or else that one model is far more plausible than another one. 

Armed with the above quantities, we propose the following decision tree when several model architectures are to be compared, with all steps illustrated in Figure\,\ref{fig:decision}. 
\begin{itemize}
    \item First, a qualitative assessment of $p(\mathcal{M})$ is necessary to decide arbitrarily which model architectures are plausible to start with. Plausible model architectures are considered realistic representations of galaxies that depend the least possible on the exact set of tracers used for constraints. The probability $p(\mathcal{M})$ typically involves parameters that cannot be varied, including the overall model architecture itself (e.g., few sectors versus\ LOC). It is worth noting that implausible models may actually lead to accurate predictions for some simple galaxy parameters (e.g., SFR, ionized gas mass).
    \item Second, it is important to evaluate the likelihood of the prior space to generate the data, which is encompassed by the marginal likelihood, $p(\bm{O}|\mathcal{M})$. Large values imply that the model architecture is complex enough given the data and that the prior space is likely to generate the data. Inversely, low values imply either that the model architecture is too complex given the data -- but may be fine if more data become available -- and/or that the prior space is simply not likely to generate the data. The combination of the first and second step corresponds to the Bayes Factor described in Equation\,\ref{eq:bf}.
  \item Finally, plausible, complex enough, model architectures whose prior space is likely to generate the data can be selected based on their predictive diagnostics. The $PPP$ can be used to distinguish between overfitting, ideal fitting, or underfitting. Model architectures selected so far but resulting in underfits correspond to well-adapted architecture, with some issues regarding for instance the data or ill-defined parameters. On the other hand, overfitting is not a problem per se since it merely reflects relatively weak constraints (or overestimated observational uncertainties) and our approach is committed to the most realistic models possible. With such overfits, the resulting posterior distribution may be particularly wide, or even identical to the prior distribution, but the predictions should remain reliable enough to interpret.      In other words, overfitting may describe a realistic model that simply needs more observables to constrain better.  
    \end{itemize}
    \noindent We emphasize that $PPP$ is the only metrics in the sequence that can be interpreted in an absolute way. The final models passing these steps may then be directly considered or even averaged. We also consider another metric for convenience in some plots of this study, which is the fraction of posterior draws matching the observed values within $3\sigma$.

\section{Application to ECO star-forming galaxy sample}\label{sec:application}

\subsection{Observations and sample}\label{sec:sample}

\begin{figure*}
\centering
   \includegraphics[width=18cm]{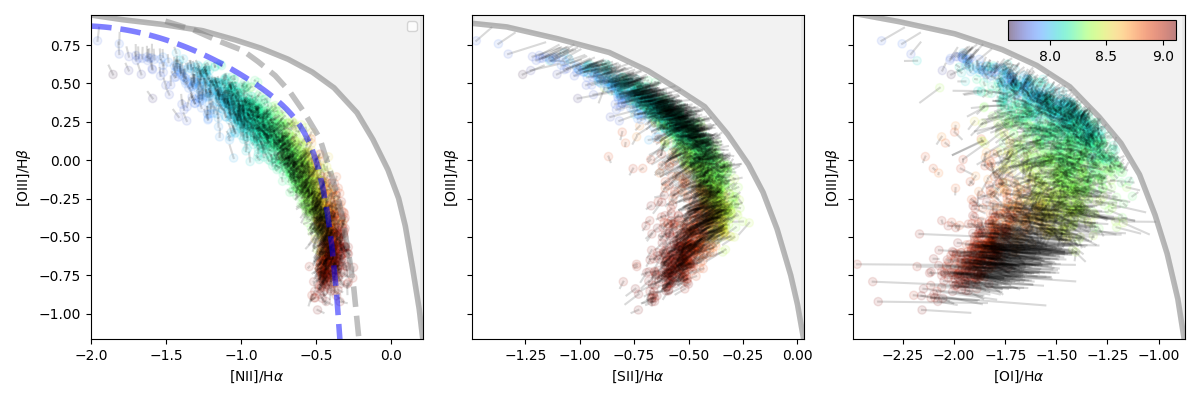}
   \includegraphics[width=18cm]{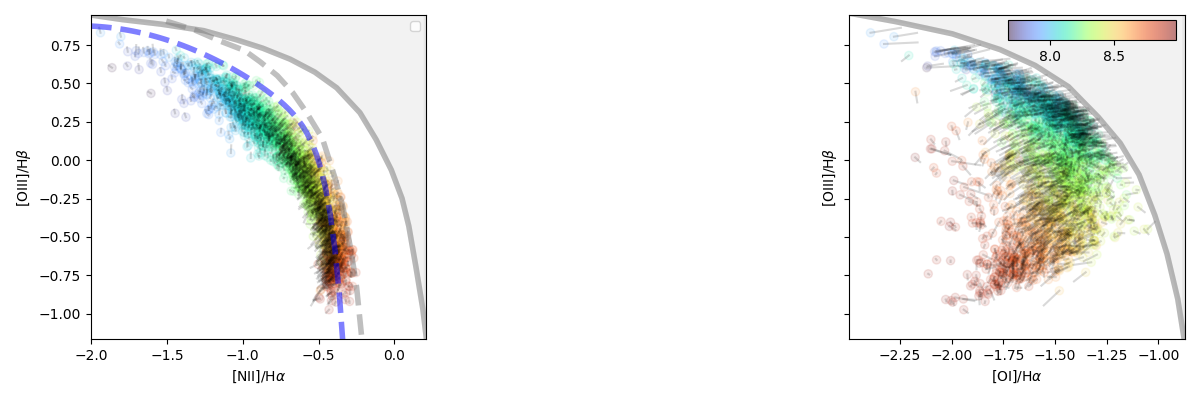}
   \caption{Excitation diagrams for the inference run with (top) and without (bottom) [S\2] lines. The color points show the modeled values (see Sect.\,\ref{sec:ecodecision}), with the color scaling with the metallicity. The solid gray curves show the extreme starburst delimiting line from \cite{Kewley2001a}, while the dashed gray curve is from \cite{Kauffmann2003a}, and the dashed blue curve is from \cite{Stasinska2006a}. The thin black lines shows the distance between the observed and modeled ratios. For [N\2]/H$\alpha$ we use the specific prescription from \citetalias{Polimera2022a} using both [N\2] lines with a scaling factor. }
   \label{fig:bpt}
\end{figure*}

We use data from the Environmental COntext (ECO) catalog DR3 \citep{Moffett2015a,Hutchens2023a}, which is a volume-limited data set in the northern spring sky spanning a recession velocity range of $3000 < cz\ [{\rm km\ s}^{-1}] < 7000$ (where $cz$ is corrected for Local Group motion and is based on group-averaged $cz$ values to minimize peculiar velocities). Being volume-limited, the sample mostly comprises low-mass dwarf galaxies. ECO has been crossmatched with SDSS spectroscopic observations by \cite[][hereafter P22]{Polimera2022a} and Polimera et al.\ (in prep.). We use the MPA-JHU catalog for the line flux measurements \citep{Tremonti2004a}, with the internal extinction corrections based on the Balmer decrement method calculated in \citetalias{Polimera2022a}. The catalog was filtered in order to use reliable detections with a signal-to-noise ratio (S/N) greater than $5$ for the strong lines H$\alpha$, H$\beta$, [O\1]\,$\lambda6300$, [O\3]\,$\lambda5007$, [N\2]\,$\lambda6548$, [N\2]\,$\lambda6584$, [S\2]\,$\lambda6717$, and [S\2]\,$\lambda6731$ (see \citetalias{Polimera2022a}).  

We select only the subset of star-forming galaxies, relying on excitation diagrams using [O\3]/H$\beta$ and ([N\2], [S\2], [O\1])/H$\alpha$ (see \citetalias{Polimera2022a} and Fig.\,\ref{fig:bpt}). Specifically we use the ``definite-SF'' category in \citetalias{Polimera2022a}. This category will be revised in Polimera et al.\ (in prep.) based on the demarcation line of \cite{Stasinska2006a}, and we have verified that the trends presented in this study remain unchanged depending on which demarcation line is used (See App.\,\ref{sec:sfagn}). We note that the S/N requirement for all lines, and especially on [O\1]\,$\lambda6300$, is meant to be able to help distinguish between star-formation and AGN activity but produces a somewhat biased toward brighter star-forming galaxies. 

The final sample is therefore almost volume-limited and comprises $2052$ galaxies. Ancillary data is available for ECO, of which we use in particular the SFR (derived from GALEX and WISE including machine-learning UV magnitude predictions for half of the sample withough deep near-UV data; \citealt{Carr2024a} and Polimera et al.\ in prep.) and the stellar mass \citep{Hutchens2023a}.

\subsection{Database of 1D models}\label{sec:grid}

\begin{figure*}
\centering
   \includegraphics[width=18cm]{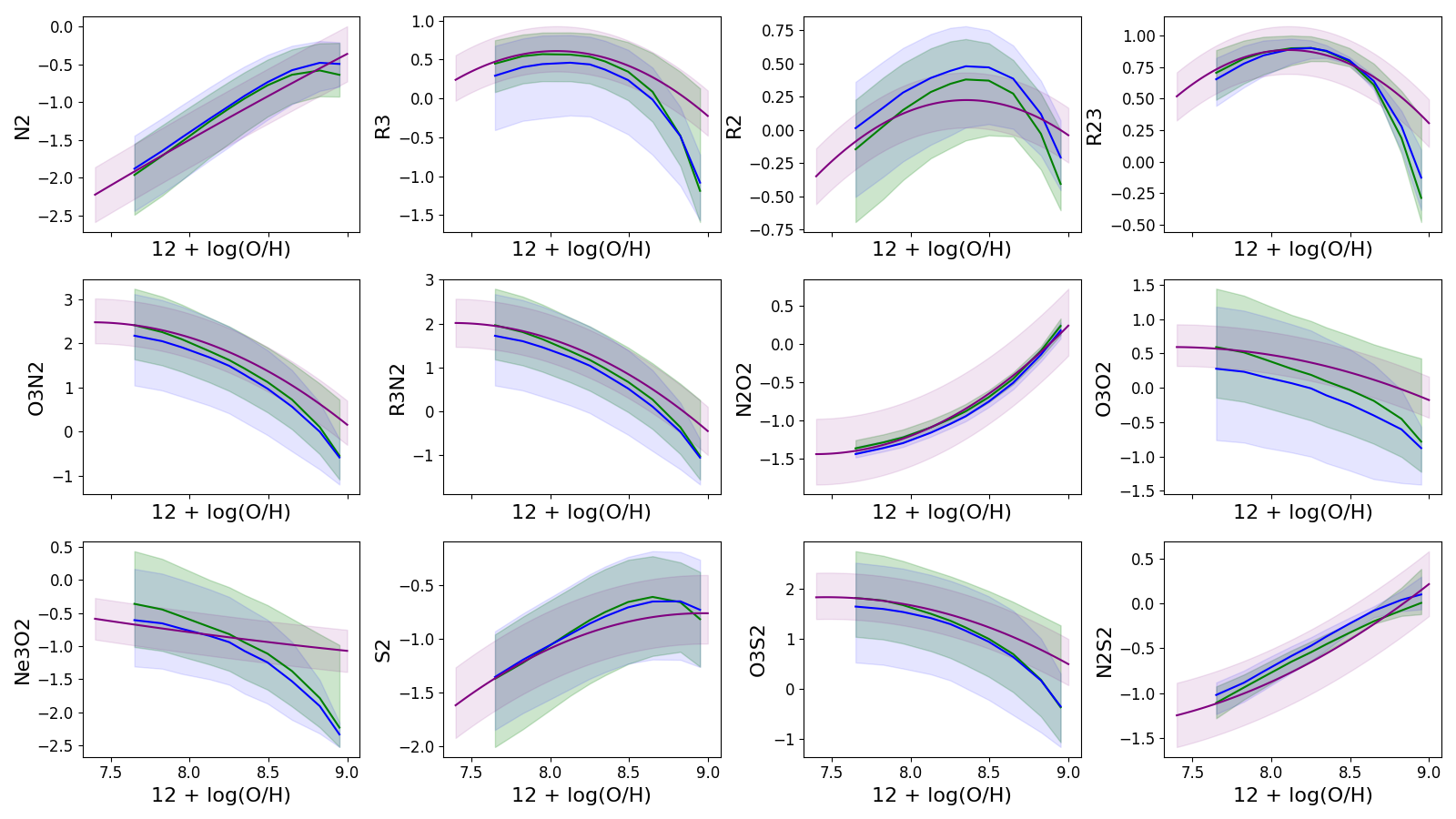}
   \caption{Empirical line ratio diagnostics as a function of metallicity for the model grid (the single 1D model is in blue and the LOC distribution is in green), compared to the calibration from \cite{Garg2024a} in purple. The shaded areas correspond to the range of physical conditions used in our photoionization grid (see text). }
   \label{fig:Zcalib}
\end{figure*}

Our modeling approach considers statitical distributions of 1D models to describe the integrated emission of galaxies (Sect.\,\ref{sec:framework}). We adopt the methodology from \cite{Richardson2022a} for the 1D photoionization models in this work. The stellar SEDs originate from the BPASS v2.0 (\citealt{Eldridge2017a}) models using an instantaneous star-formation burst (i.e., simple stellar population). The post-starburst ages span $1-25$\,Myr, which ensures that multiple Wolf-Rayet stages are captured (see \citealt{DAgostino2019a}). In actuality, the post-starburst age essentially accounts for the hardness of the radiation field since any given star-forming region realistically has a mixture of stellar ages. The stellar metallicities range from $0.05$\,Z$_{\odot}$ to $2.0$\,Z$_{\odot}$. Our model grid extends to $3.0$\,Z$_{\odot}$, but BPASS models are unavailable at these metallicities, so we substitute $2.0$\,Z$_{\odot}$ models in the cases where $Z > 2.0$\,Z$_{\odot}$. 

The abundance scalings in the cloud are taken from the Galactic Concordance Abundances described in \cite{Nicholls2017a} where the solar standard is defined as $12+\log({\rm O/H})=8.76$. We use the depletion patterns described in the appendix of \cite{Richardson2022a} for a fixed depletion strength of $F_* = 0.45$. The model grid uses the parameter $Z/{\rm Z}_{\odot}$ for metallicity, which refers to the abundances in the cloud prior to grain depletion. Therefore, one needs to deplete the oxygen abundance by $-0.11$\,dex to obtain ``gas-phase abundances'' in terms of $12+\log({\rm O/H})$. We assume a grain composition like the Orion Nebula \citep{Baldwin1991a}, in addition to including polycyclic aromatic hydrocarbons \citep{Abel2008a}, and use a D/G ratio scaled with metallicity according to the empirical relation from \cite{RemyRuyer2014a}. The hydrogen density at the face of the gas cloud, log~$n_{\text{H}}$ varies from $0.5$ to $4.0$ in $0.5$\,dex increments, while the ionization parameter $U$, also defined at the ionized face, varies from $-4.5$ to $-0.5$ in $0.25$\,dex increments. The models are run until an electron fraction of $n_e/n_{\rm H} = 0.01$.

The integrated distributions (LOC) models are drawn from this grid of 1D models. We briefly describe tests using other photoionization codes in Section\,\ref{sec:bimodality}. Our grid includes a potential radiative component powered by an AGN, but for the present star-forming sample we select only a subgrid with an "AGN fraction" (i.e., fraction of ionizing radiation due to an AGN) of $f_{\rm AGN}=0$. We defer the study of galaxies with significant or dominant AGN fraction but tests have been performed to verify that the AGN fraction for the present sample, if let free, never reaches above $8$\%  (see \citealt{Polimera2022a}) with most galaxies showing $f_{\rm AGN}<4\%$.

Before inferring the metallicity and other parameters, we first compare the metallicity from the grid of single 1D models against empirical line ratio diagnostics from \cite{Garg2024a}\footnote{Contrary to what is indicated in \cite{Garg2024a}, we do use the [O\2] doublet sum for N2O2 and we do consider a sum of logarithms for N2S2.}. The metallicity from the grid controls the elemental abundances that, in turn, are used to compute the radiative transfer within Cloudy. For this comparison, we restrict the parameter ranges (average for LOC) to $12+\log({\rm O/H})=[7.4,9.0]$, $U=[-3.5, -2]$, and $n=[1, 2]$ to match the ranges used for calibration. For the LOC distributions, we consider boundaries for $U$ and $n$ that are typically found in the present study ($[-4, -2]$) and $[1, 3]$ respectively) and ensure that the average values remain within the calibration ranges. We find good agreements overall across the considered metallicity range, especially with the N2O2, N2S2, N2, S2, R3N2, and O3N2 diagnostics (Fig.\,\ref{fig:Zcalib}). This implies that the metallicity parameter in the grid corresponds, to first order, to the metallicity obtained from the empirical diagnostics. 

Differences likely stem from the assumed N/O abundance ratio prescription in the various calibrations (see specific discussion in \citealt{Garg2022a}). Any deviation will result in important biases if we wish to compare the metallicity from the grid (e.g., the value inferred through modeling observations) to the metallicity from empirical calibrations and may highlight specific model assumptions regarding, for instance, the depletion pattern of some elements as a function of metallicity. 

The best agreement between our models and empirical diagnostics is found for N2O2, but the [O\2] doublet is unfortunately difficult to measure and available only for a small number of ECO sources and with large uncertainties. All other relatively reliable empirical diagnostics show some kind of deviations compared to our models. Therefore, we keep in mind in the following that the metallicity inferred from our models may deviate somewhat from that obtained with empirical diagnostics.

\subsection{Model architectures for ECO star-forming galaxies}

\subsubsection{Relevant architectures}\label{sec:relevant}

Based on the a priori model probability $p(\mathcal{M})$ criterion (Sect.\,\ref{sec:metrics}), LOC models (using either power-law or normal distributions) are preferred to multicomponent ``$x$C$y$S'' models due to the evidence of heterogeneity of ISM and energetic source properties (see introduction). Building upon the recent modeling effort from \cite{Ramambason2024a}, we present here the first application of LOC models with the age and $Z$ following a statistical distribution, in addition to $n$ and $U$. For the integration boundaries of LOC models ($p_{\rm min,max}$), MULTIGRIS considers by default the minimum and maximum values in the grid for each parameter, but free boundaries are considered in the present study.

Overall, many different architectures have been investigated with the number of random variables ranging from $5$ to $\gtrsim20$ (Table\,\ref{tab:rvs}) and we will only focus on a few model architectures afterward. For simplicity, we present here the results assuming power-law distributions only. The reasoning is that, on first order, narrow normal distributions may be approximated by single parameter values while broad normal distributions may be approximated by flat power laws. Although it would be interesting to thoroughly test various distributions, we keep in mind that we may not be able to afford fine-tuning in the model architecture (e.g., testing a power law versus\ normal distributions) considering the various other caveats (e.g., projection effects, 1D model grid hypotheses). 

We consider ``$x$C$y$S'' models strictly for comparison and continuity with previous works, despite the fact that these models are considered comparatively less realistic because they assume that the excitation sources or the ISM are fully described by one or two components typically, i.e., by a single parameter value (e.g., single density) or by a combination of two parameter values (e.g., two components with a single density for each).

  \begin{table}
  \caption{Subset of tested model architectures.}\label{tab:rvs}
\begin{tabular}{lllll}
\hline
\hline
 &  1C1S & 2C1S &  LOC $U$/$n$   & LOC $U/n/{\rm age}/Z$     \\
  \hline
& $\delta_{\rm age}$  & $\delta_{\rm age1}$, $\delta_{\rm age2}$ & $\delta_{\rm age}$  &  $\alpha_{\rm age}$, age$_{\rm min,max}$ \\
&   $\delta_U$ & $\delta_{U1}$, $\delta_{U2}$    &  $\alpha_U$, $U_{\rm min,max}$ & $\alpha_U$, $U_{\rm min,max}$ \\
& $\delta_n$  & $\delta_{n1}$, $\delta_{n2}$   & $\alpha_n$, $n_{\rm min,max}$ & $\alpha_n$, $n_{\rm min,max}$   \\
&   $\delta_Z$ & $\delta_{Z1}$, $\delta_{Z2}$  & $\delta_Z$  & $\alpha_Z$, $Z_{\rm min,max}$ \\
  \hline
RVs &  4 & 8 & 8  & 12 \\ 
\hline
\end{tabular}\\\footnotesize{
Notes -- $\alpha$, $\delta$, as well as minimum and maximum values refer to the distribution hyperparameters (see Eqns.\,\ref{eq:alpha} and \ref{eq:delta}). RVs indicates the raw number of random variables (ignoring potential correlations). }
\end{table}

\subsubsection{Model selection}\label{sec:ecodecision}

Based on the marginal likelihood $p(\bm{O}|\mathcal{M})$, the single component model (1C1S) is not favored compared to the two component models (1C2S or 2C1S) because the prior space is too simple and therefore not likely to generate the data. LOC models, on the other hand, may quickly become too complex given the data. As mentioned earlier, this is not a problem per se, and we aim to compare and select the best model architectures that are deemed realistic enough. 

Based on the $PPP$, we find that 1C2S or 2C1S models perform quite well, and the only reason we do not fully consider them for interpretation is that they were not selected initially as realistic models based on the a priori probability of the model $p(\mathcal{M})$ (Sect.\,\ref{sec:relevant}). The reason why these models perform so well is due to the parameter flexibility: the parameters for the few (or single) considered components are independent, a single value per component is fine-tuned to match the observations, and the value is not necessarily expected to correspond to a physically meaningful region of the galaxy. 

Among LOC models, the best models are those with free parameter boundaries ($p_{\rm min,max}$) for the integration. Even though models with free boundaries involve a larger number of random variables (Table\,\ref{tab:rvs}), the LOC models with free boundaries do not indicate significant overfitting ($PPP<0.5$) and show only slightly lower marginal likelihoods compared to the models with fixed (minimum and maximum in the grid) boundaries due to the expanded parameter prior space. Our final model architectures are LOC models with power-law distributions and free parameter boundaries.

It should be emphasized that the match between the LOC models and observations is driven simultaneously by the topology assumptions (parameters describing such combinations) and by the inherent 1D model database (emission line predictions from a set of parameters such as metallicity, ionization parameter, etc.). We choose to design the most adequate photoionization grid possible and focus on topology improvements, but we keep in mind that our results are strongly dependent on the reference grid and that other conclusions could be reached with different prescriptions, e.g., for the radiation sources, or the ISM composition (see, e.g., \citealt{Lecroq2024a}). We also remind that the distributions we implement do not correspond to the entire ISM of galaxies but are biased toward the emitting components and, as such, are driven by H\2\ region properties in this star-forming galaxy sample. 

\section{Results}\label{sec:results}

An illustration of the inference results obtained for individual galaxies is presented in App.\,\ref{sec:singleresults}. By default, we consider in the following probability density functions (PDFs) for the entire sample. For this, we do not simply select the mean of each parameter for each object but concatenate the draws for all the objects in order to keep the information contained within the confidence intervals.

\subsection{Influence of the line set}\label{sec:lineset}

\begin{figure}
\centering
   \includegraphics[width=9cm]{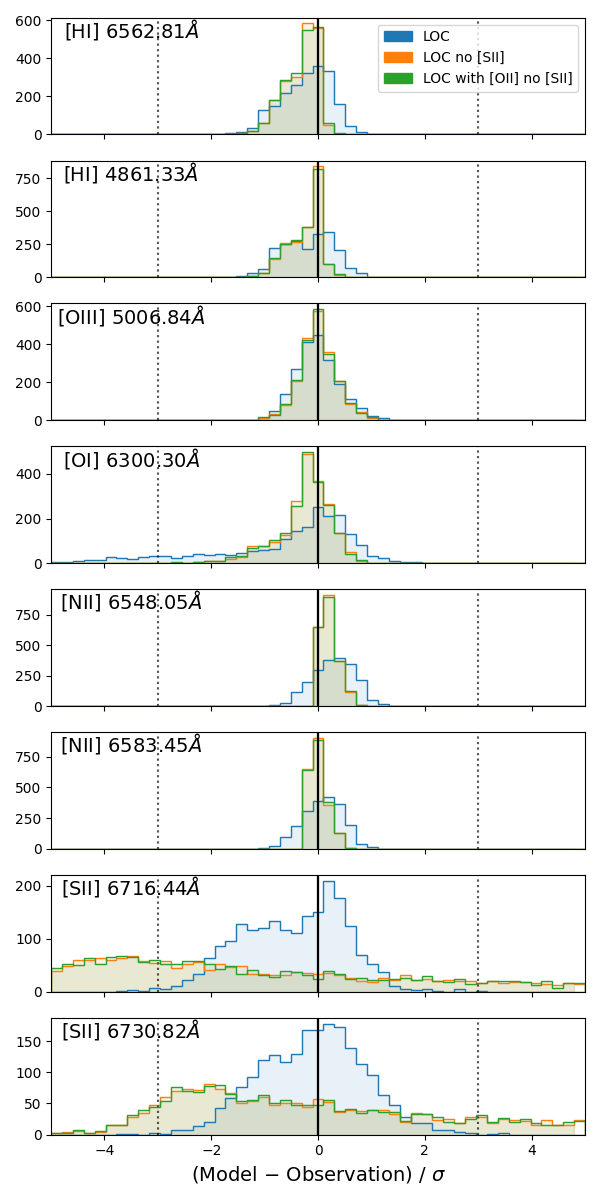}
   \caption{Histogram comparison between predicted and observed fluxes scaled by the observed errorbar for the entire sample. The vertical lines delimit the agreement within $3\sigma$. }
   \label{fig:fluxmatch}
 \end{figure}

\begin{figure}
\centering
   \includegraphics[width=9cm]{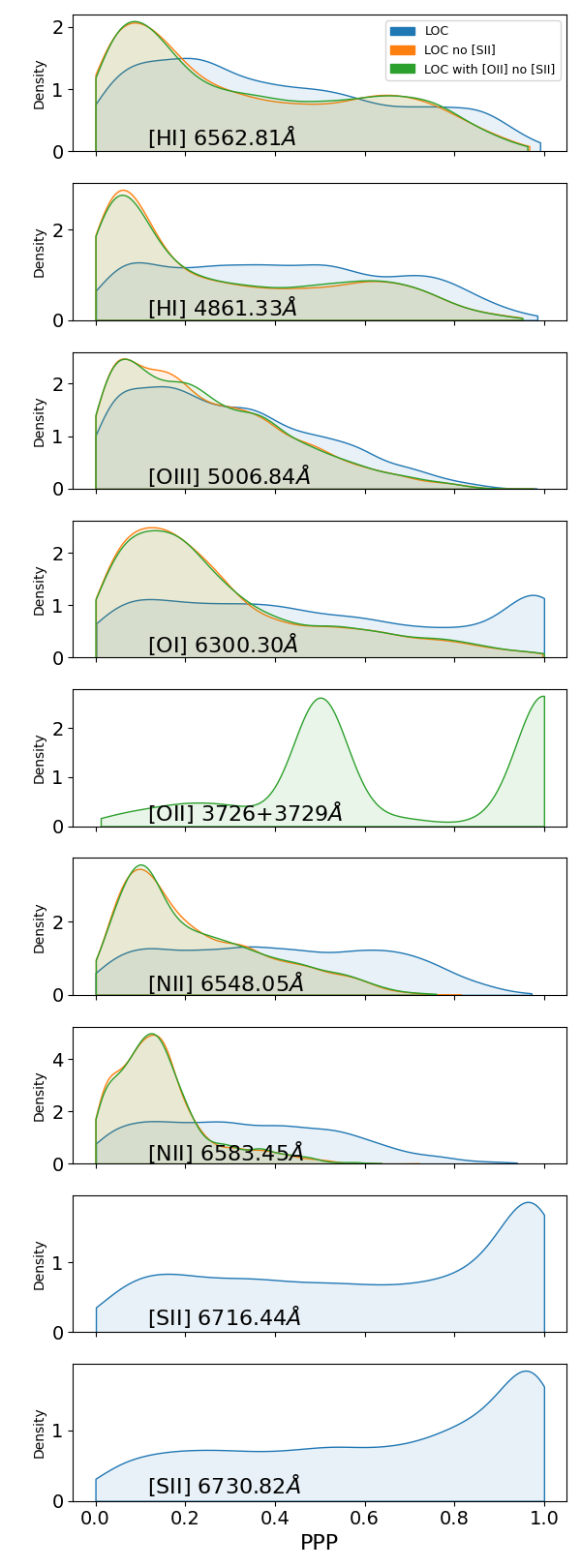}
   \caption{Posterior predictive $p$-value ($PPP$) for each line.}
   \label{fig:performance_lines}
 \end{figure}

Figure\,\ref{fig:fluxmatch} shows that the predicted line fluxes agree within $\approx2\sigma$ for all galaxies in the sample with the LOC architecture using all available lines. Looking at the specific posterior predictive $p$-value ($PPP$) for each line for the 1C1S and LOC architectures using all available lines (Fig.\,\ref{fig:performance_lines}), we find that underfitting is largest for [S\2], [O\1], and [N\2], all being slightly overpredicted by the model. Motivated by the underfitting of [S\2] lines as well as potential deviations between empirical metallicity diagnostics involving [S\2] and the metallicity from the grid (Sect.\,\ref{sec:grid}), we ran the inference without the [S\2] lines as constraints. We also considered runs with the [O\2] doublet sum, despite with poor signal-to-noise ratio, instead of the [S\2] lines. 

Figure\,\ref{fig:performance} shows that LOC models ignoring the [S\2] lines perform much better. This remains true even using [O\2] instead of [S\2]. Furthermore, Figure\,\ref{fig:performance_lines} shows that [N\2] and [O\1] are dramatically better matched ($PPP\leq0.5$) in runs ignoring [S\2] (see also Fig.\,\ref{fig:bpt}). In the following, we will consider runs that include or do not include [S\2] or [O\2] to study the impact of the line set on our results. The overprediction of [S\2] in the models may be due to the 1D model grid assumptions (e.g., need to account for sulfur depletion, need for more refined stellar atmospheres for the relevant energy range; Sect.\,\ref{sec:grid}) and/or to systematic effects in the line measurement available in the SDSS catalogs (see  discussion in \citealt{Polimera2022a}) that may be due to the difficulty in removing nearby telluric features. We also have verified that ignoring the [O\1] line for the inference of the entire sample does not significantly modify the metallicity determination.

\subsection{Performance of single 1D models (1C1S) versus LOC}\label{sec:perf}

\begin{figure*}
\centering
   \includegraphics[width=9cm]{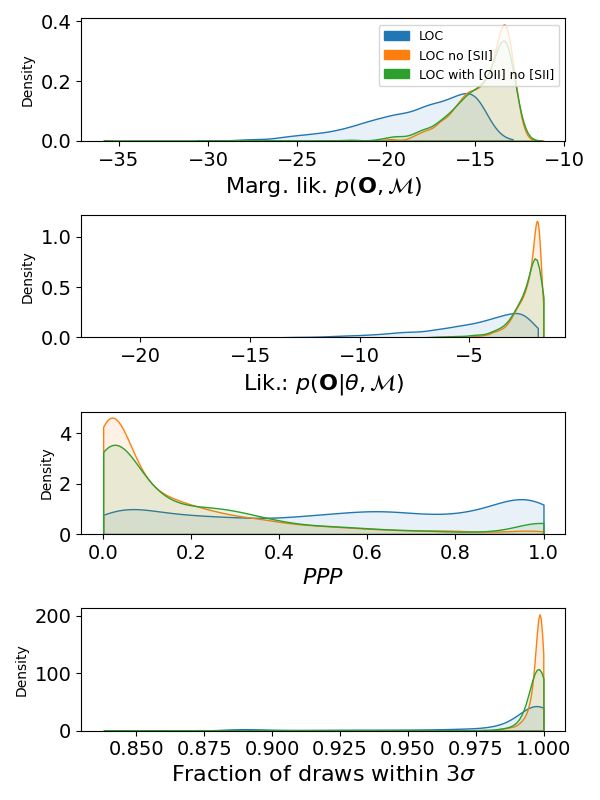}
   \includegraphics[width=9cm]{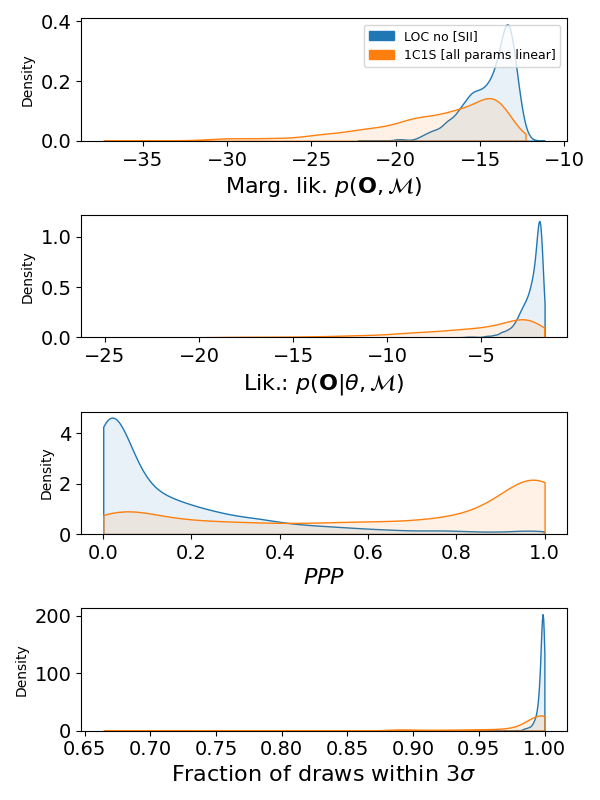}
   \caption{Performance metrics for the entire sample. From top to bottom: the marginal likelihood and the likelihood (evidence), the absolute posterior predictive $p$ value ($PPP$), and the fraction of posterior draws matching the observations within $3\sigma$.}
   \label{fig:performance}
\end{figure*}

Here we wish to compare the -- often used -- single 1D model approach (1C1S) to the LOC one. For this comparison we are therefore interested in biases specifically due to the model architecture. For this test the LOC boundaries ($p_{\rm min,max}$) are not linearly interpolated but the slope ($\alpha_p$) is continuously sampled. Hence we consider 1C1S with all parameters linearly interpolated instead of nearest neighbor for a fairer comparison, keeping in mind that LOC models would perform even better through boundary interpolation between models. 
  
Figure\,\ref{fig:performance} shows that, even considering the best possible 1C1S models (with linear interpolation for all parameter), LOC models always perform better in all metrics. While we show for reference the results for the inference ignoring [S\2] in Figure\,\ref{fig:performance}, the same results hold for all inference runs, and also hold for tests with only one or two parameters following statistical distributions instead of single values.

The fact that LOC models globally outperform 1C1S models strengthens the hypothesis that physical conditions are not uniform within the galaxies of the sample, and consequently that the integrated tracers we observe do contain useful information about the distribution of matter and radiation sources. Since our tests show that $PPP$ does not improve significantly beyond (any) two parameters being distributed as power laws, we conclude that this is most likely the minimum amount of complexity dictated by the set of tracers we use. However, we keep in mind that physical motivations exist for all parameters to follow some kind of statistical distributions and that LOC models are considered more likely a priori (independently on observed tracers, i.e., with a larger $p(\mathcal{M})$ value) than multicomponent ``$x$C$y$S'' models.

\subsection{Physical meaningfulness of LOC average and single 1D model values}\label{sec:locvs1d}

While we show in Section\,\ref{sec:perf} that the LOC approach outperforms the single 1D model approach, potential biases on physical parameter determinations due to the chosen approach need to be addressed. Output physical parameters using single 1D models are often interpreted as some kind of average quantities and we put this assumption to the test by comparing the single parameter values of 1C1S models to the average quantities from LOC models (see Eqn.\,\ref{eq:avg}). 

\begin{figure*}
\centering
   \includegraphics[width=18cm]{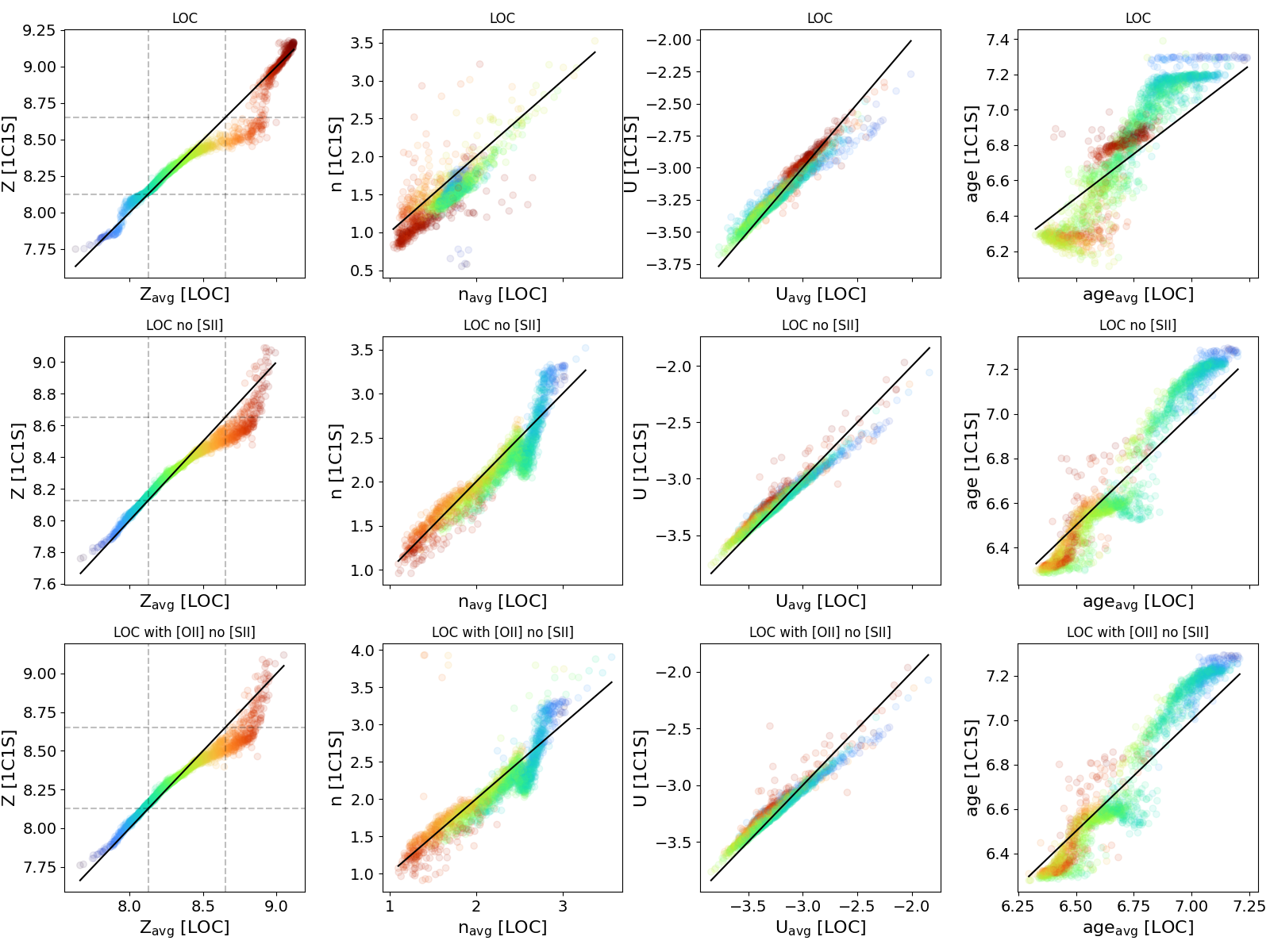}
   \caption{Comparison of single 1D models (1C1S) versus\ LOC averages. The color scales with the metallicity parameter. From top to bottom we show the results for the runs ignoring [O\2], ignoring [S\2] and [O\2], and ignoring [S\2] (Sect.\,\ref{sec:lineset}). The dotted lines in the leftmost plots indicate the $0.3$\,Z$_\odot$ and Z$_\odot$ values. }
   \label{fig:averages}
\end{figure*}

Globally the 1C1S values are always compatible with the LOC averages within a factor of a few (Fig.\,\ref{fig:averages}). We keep in mind that for all parameters, 1C1S or LOC averages never reach the grid minimum or maximum values and that there is no edge effect.
We may either consider the 1C1S approach as some kind of reference because it is often used and it is therefore reassuring that the LOC average value matches the 1C1S value, or we may also consider the LOC approach to be more realistic and viable and it is reassuring that 1C1S models provide values that do not deviate significantly. However, it should be restated that the 1C1S models globally perform less well than LOC and that the biases we identify are interpreted as biases due to the single 1D model hypothesis (in other words, we are not trying to validate the average LOC quantities). 

Looking more closely, we find some small deviations. For the metallicity, there is little bias for very low- and very high-$Z$ but there is a ``kink'' around solar metallicities with the 1C1S $Z$ somewhat lower than the LOC average $Z$, especially for the run that includes the [S\2] lines. We find that this bias is not due to one single specific parameter in the LOC models, but to all of the parameters in aggregate. For instance, the relatively wider range of $Z$ (boundaries) around solar metallicities could explain in part this bias but also the consideration of ranges (LOC) for $U$ or age that results in a range of $Z$ because of the dependency between both $U$ and age with $Z$ (see discussion in Sect.\,\ref{sec:correlations}).   

The age (i.e., spectral hardness of the input radiation field) parameter shows significant biases, with the 1C1S age being overestimated in the most metal-poor galaxies (blue points in Fig.\,\ref{fig:averages}) and being underestimated in slightly subsolar to solar metallicity galaxies (orange points). There is almost no bias for the highest-$Z$ galaxies (red points). 

For the ionization parameter $U$, there is no clear bias. For the density $n$, there is also no clear bias apart from globally slightly lower values with 1C1S compared to LOC averages. This is likely due to the relatively poorer performance of 1C1S (all parameters linearly interpolated) including and especially with [S\2] lines. 

Globally, we notice that there seems to be a special regime around solar metallicity corresponding to a turnover in $U$ and age that may lead to significant biases using single 1D model results. As a function of metallicity, there is a clear trend for the age (and therefore hardness) to decrease between $20$\,Myr and $5$\,Myr until solar metallicity and increase again. Similarly, as a function of metallicity, there is a clear trend for the ionization parameter to decrease (until slightly subsolar metallicities) and increase again. Thus, we definitely see a relationship between age, $Z$, and $U$ (see also Sect.\,\ref{sec:correlations}). We note that these turnovers are not driven by the grid minimal or maximal values for each parameter in the grid.

In summary, biases may exist using single 1D models that may affect the interpretation of the inferred parameters or other, related, parameters. The physical parameters tackled in this study are related to H\2\ regions, and their average values do not depend significantly on the ISM topology. However, stronger biases may exist for other specific galaxy parameters (e.g., H$_2$ masses, escape fraction of ionizing photons, etc.; \citealt{Ramambason2024a}).

\subsection{Inferring internal parameter distributions within galaxies}\label{sec:parametersandcorrelations}

In the following we examine the PDF of the average parameter values within galaxies $Z_{\rm avg}, U_{\rm avg}, n_{\rm avg}, {\rm age}_{\rm avg}$ (not to be confused with the average value across the sample). For some parameters (e.g., $Z_{\rm avg}$ and $U_{\rm avg}$), the PDFs for individual objects are much narrower than the sample PDF, implying that the latter describes well the different properties of the studied galaxies (Figs.\,\ref{fig:hyperaverages} and\ \ref{fig:params_indiv}). For the other parameters (in particular $n_{\rm avg}$), the PDFs for individual objects are identical on first order and the overall sample PDF thus reflects a common PDF, valid for all galaxies. 

The PDFs of the power-law slopes ($\alpha$) are similar on first order for all galaxies for any given physical parameter, which may indicate a universal origin of the distribution but could also reflect the difficulty in constraining $\alpha$ from the observed tracers used for inference. Considering the PDFs from Figure\,\ref{fig:params_indiv}, we propose that small variations of $\alpha$ for a given galaxy will lead to significant variations of the boundaries, and that the observed tracers mostly constrain the average physical parameter value. Small variations do exist, however, for all hyperparameters, from galaxy to galaxy and from the prior distribution, and we investigate them in the following.

\subsubsection{Hyperparameters}\label{sec:hyperparameters}

\begin{figure*}
\centering
   \includegraphics[width=9cm]{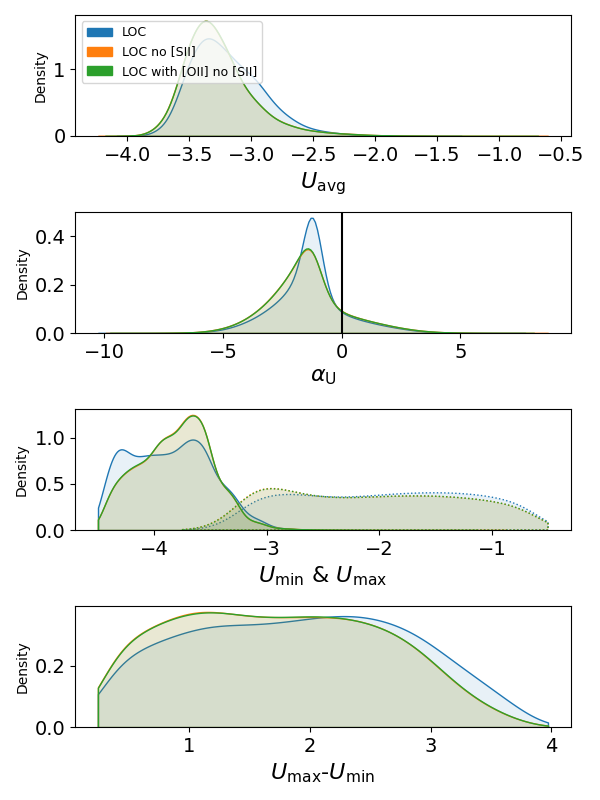}
   \includegraphics[width=9cm]{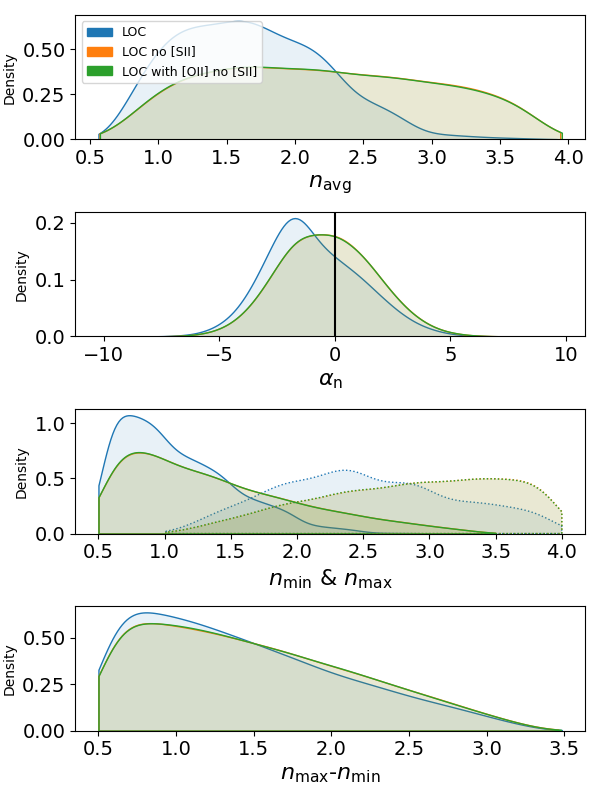}
   \includegraphics[width=9cm]{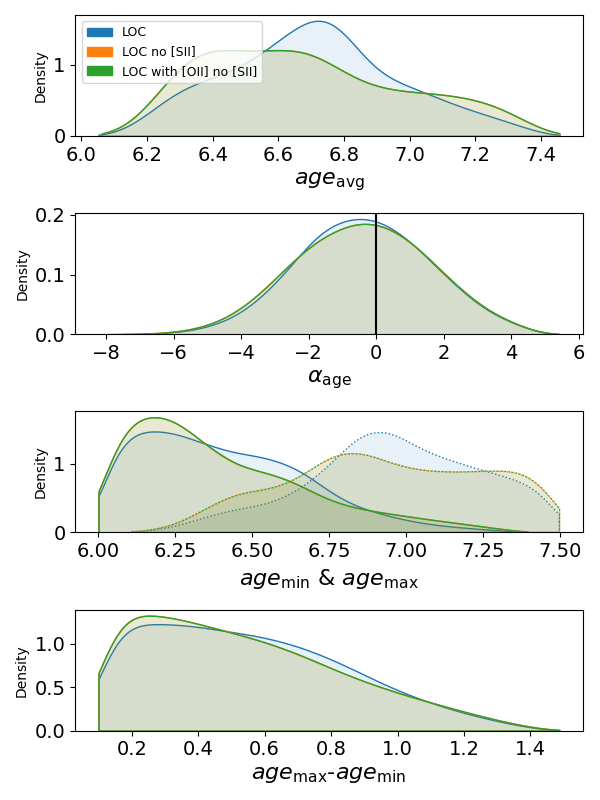}
   \includegraphics[width=9cm]{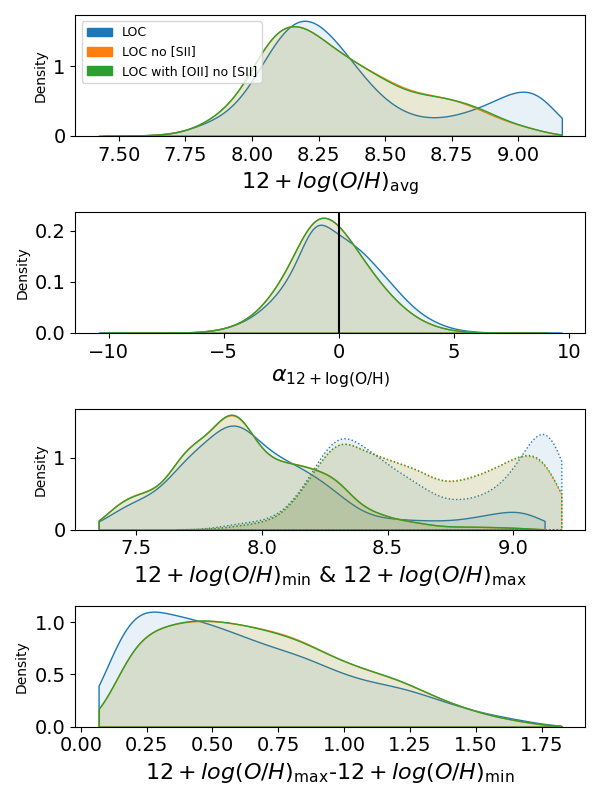}
   \caption{Hyperparameter and average values. }
   \label{fig:hyperaverages}
\end{figure*}

The hyperparameter $\alpha$ (slope of the power-law distribution) reflects the relative proportions of emitting components with given physical properties. The slopes for age, density, and metallicity are close to $0$ (Fig.\,\ref{fig:hyperaverages}), hinting that for most galaxies the emission is not significantly dominated by a given dense versus diffuse, old versus young, or high versus low metallicity.
For most galaxies, however, the emission is dominated by relatively low excitation components ($\alpha_U<0$), with $\log U_{\rm avg}$ fairly peaked around $\approx-3.2$. The low-excitation components, that contribute most to the total emission of most galaxies, show a narrow range of $\log U_{\rm min}$ between $-4.5$ and $-3$. Inversely, the higher-excitation components show a wide range of $\log U_{\rm max}$ centered around $-2$. 

The distribution of ${\rm age}_{\rm avg}$ peaks around $5$\,Myr. The distribution of the lower and upper boundaries peak around ${\rm age}_{\rm min}\approx2$ and ${\rm age}_{\rm max}\approx10$\,Myr respectively, but the upper boundary extends to the Wolf-Rayet phase at $\approx20$\,Myr, which is the hardest radiation field in the grid \citep{DAgostino2019a}.

The distribution of $n_{\rm avg}$ almost spans the entire parameter space, except if [S\2] lines are used as constraints. While it is expected that [S\2] lines help constrain the density (e.g., \citealt{Osterbrock2006a}), we note that these lines may also cause some biases (Sect.\,\ref{sec:lineset}).  

The distribution of $Z_{\rm avg}$ is bimodal, with a stronger bimodality for the inference run using the [S\2] lines. Most galaxies lie around $Z_{\rm avg}\sim0.3$\,Z$_\odot$ and populate the leftmost peak. The secondary peak lies around $2$\,Z$_\odot$ if [S\2] lines are used, and otherwise around solar metallicity. The distribution of the lower boundary peaks around $Z_{\rm min}\approx0.15$\,Z$_\odot$. The distribution of the upper boundary $Z_{\rm max}$ is strongly bimodal (Fig.\,\ref{fig:hyperaverages}), clearly driving the bimodality of the average metallicity. 

In individual galaxies, it must be emphasized that boundaries for all parameters are mostly well separated (Fig.\,\ref{fig:hyperaverages}), yet there is no prior to force a minimum difference between the maximum and minimum value to be considered for integration. In other words, the inference could have resulted in boundaries being equal or almost equal (i.e., being equivalent to a single 1D model) if this had been a more likely solution (see also App.\,\ref{sec:singleresults}).

\subsubsection{Correlations between physical parameters}\label{sec:correlations}

\begin{figure*}
\centering
   \includegraphics[width=6cm,trim=0 47 0 0,clip]{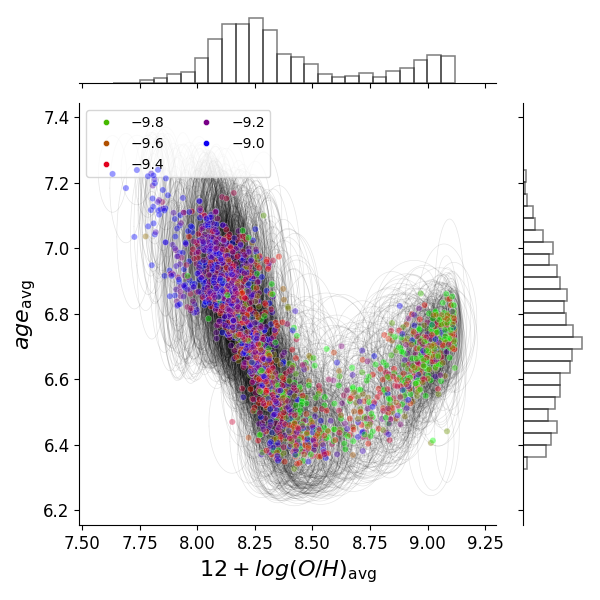}
   \includegraphics[width=6cm,trim=0 47 0 0,clip]{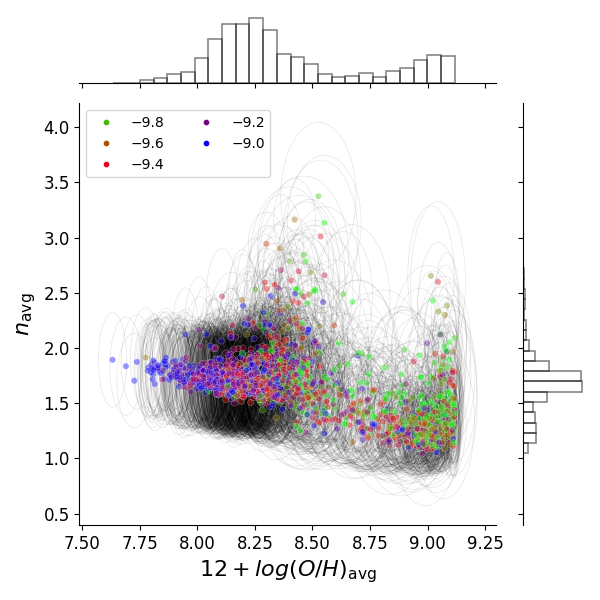}
   \includegraphics[width=6cm,trim=0 47 0 0,clip]{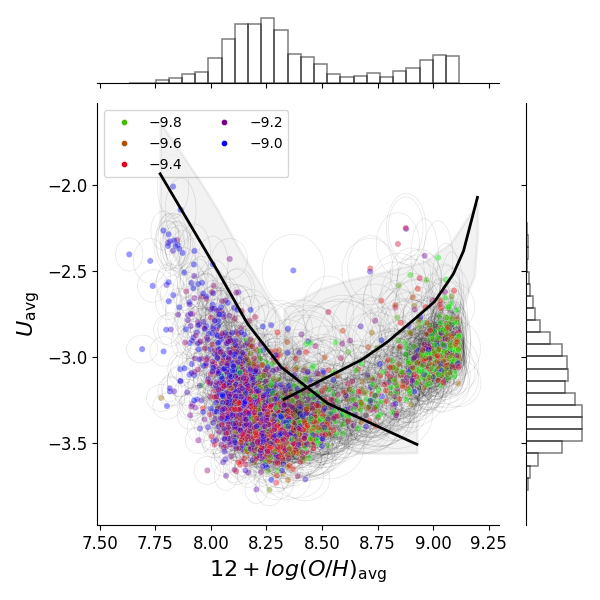}\\
   \includegraphics[width=6cm,trim=0 0 0 60,clip]{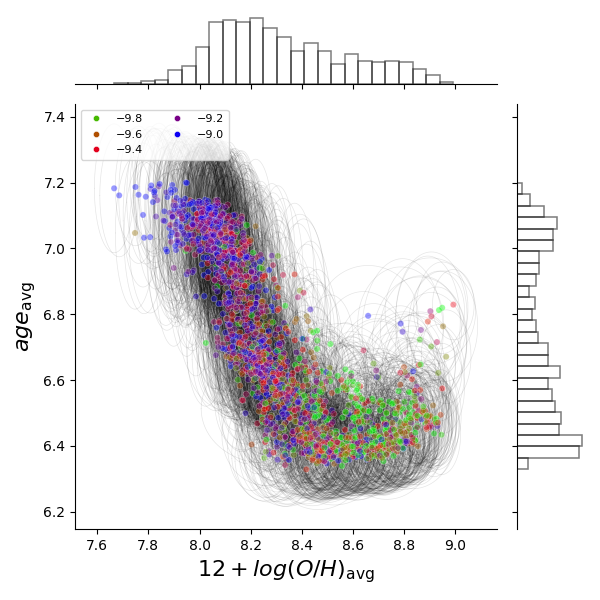}
   \includegraphics[width=6cm,trim=0 0 0 60,clip]{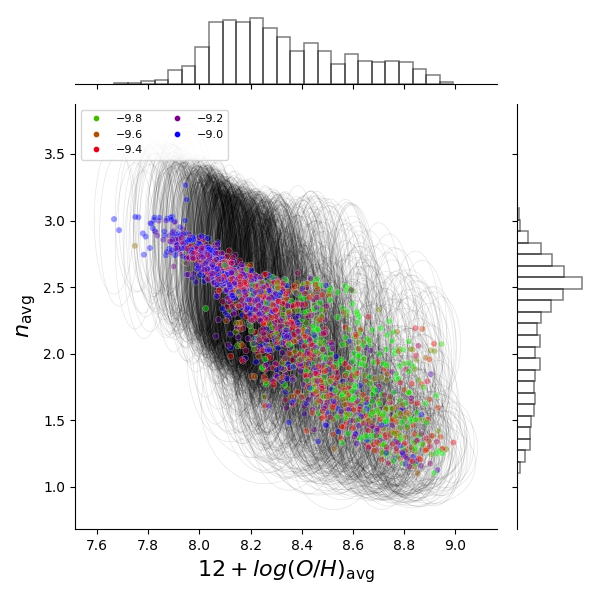}
   \includegraphics[width=6cm,trim=0 0 0 60,clip]{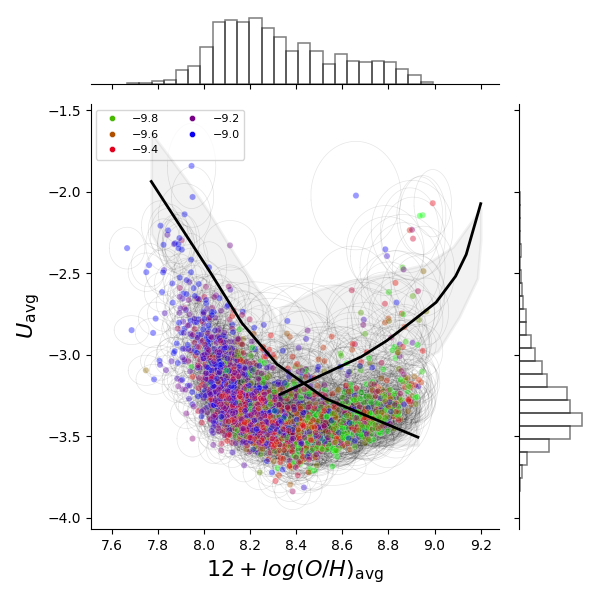}
   \caption{Correlations between physical parameters for the runs including (top) or excluding (bottom) the [S\2] lines. The color scale indicates the sSFR. For $U$ versus $Z$, the low- and high-$Z$ curves are from \cite{Kashino2019a} and \cite{Ji2022a} respectively.}
   \label{fig:correlations}
\end{figure*}

We investigate the correlation between physical parameters using their average value in each galaxy. Results are shown in Figure\,\ref{fig:correlations} for the inference runs with and without [S\2]. We note that there is no degeneracy between the parameters and that the PDFs of individual galaxies clearly prefer one solution (see example in Fig.\,\ref{fig:illustration_single_pairplot}).

There is no clear trend between ${\rm age}_{\rm avg}$ and $U_{\rm avg}$. However, we find a strong relationship between ${\rm age}_{\rm avg}$ and $Z_{\rm avg}$. The most metal-poor galaxies are characterized by ${\rm age}_{\rm avg}\approx10$\,Myr. Around slightly subsolar metallicities, ${\rm age}_{\rm avg}$ reaches down to $\approx3$\,Myr, i.e., a softer radiation field. For high-metallicity galaxies, the runs including the [S\2] lines indicate older ages around $\approx5$\,Myr and therefore intermediary hardness, while the runs ignoring [S\2] flatten around $\approx3$\,Myr. Our sample is selected based on [O\1] detection, thereby selecting relatively hard radiation fields, but other high-metallicity galaxies may actually not require such hard radiation field. The tendency for high-$Z$ sources to require a hard radiation field could also be indicative that other high-ionization process may be important (e.g., shocks).

The $U_{\rm avg}$ versus\ $Z_{\rm avg}$ correlation shows the same trend as ${\rm age}_{\rm avg}$ versus\ $Z_{\rm avg}$ but with the turn-off occurring at a lower metallicity. The slight increase of $U_{\rm avg}$ above $0.3$\,Z$_\odot$ for the runs including [S\2] is reminiscent of the result obtained in high-redshift star-forming galaxies \citep{Reddy2023a}. The average density $n_{\rm avg}$ also shows a decreasing trend with metallicity, from $\sim500$\,cm$^{-3}$ down to $\sim50$\,cm$^{-3}$, with a significantly tighter trend for the runs ignoring [S\2]. 

Although not shown, we observe the same trends for the slopes $\alpha_{\rm age}$, $\alpha_U$, and $\alpha_n$ versus\ $Z$ as for the average parameters. Since the slopes reflect the weight of regions with given physical parameters toward the integrated galaxy emission, this implies that as $Z_{\rm avg}$ decreases, there is an increasing proportion of harder stellar radiation field, high ionization parameter, and high density contributing to integrated galaxy spectrum.

\section{Discussion}\label{sec:discussion}

\subsection{Ionization parameter versus metallicity}\label{sec:relations}

There is evidence in the literature that the ionization parameter $U$ and the metallicity $Z$ are physically related but the origin of the relationship is still debated (see, e.g., \citealt{Dopita2006b,Ji2022a}). First, there is evidence that $U$ anticorrelates with $Z$ in low-metallicity galaxies until about solar metallicity (or stellar mass $\sim10^{10}$\,M$_\odot$) (\citealt{Kashino2019a,Reddy2023a}; see black curves in Fig.\,\ref{fig:correlations}). Independently, there is also evidence that $U$ correlates with $Z$ for metal-rich sources, above solar metallicity \citep{Ji2022a}. Suprisingly, there are few or no samples spanning a wide enough range of metallicities to verify whether these relations are specific to some given metallicity regimes. The ECO sample is ideal for studying this relationship because we have access to a wide range of masses and metallicities.

Our results show a relatively well-behaved relationship between $U_{\rm avg}$ and $Z_{\rm avg}$, with a steep decline followed by a smooth increase (almost nonexistent if [S\2] lines are ignored; see Fig.\,\ref{fig:correlations} bottom). Our results are in line with theoretical expectations. The steep decline could be explained by the wind-driven bubble model for H\2\ regions of \cite{Dopita2006a}, which would dominate at low-metallicity, together with the lower opacity of low-metallicity stellar atmospheres resulting in greater ionizing flux. This interperation is strengthened by models of multiple H\2\ regions within a single galaxy  \citep{Garner2025a}. It should be noted that \cite{Kashino2019a} show that, despite the strong apparent anticorrelation between $U$ and $Z$ in their low-metallicity sample, the $U$ variation depends more heavily on the specific SFR (sSFR). In summary, $U$ may be controlled by a competition between variations of $Z$ and of sSFR, with a moderate $U$ versus\ $Z$ anticorrelation at low-metallicity steepened by sSFR. 

On the other hand, the smooth increase of $U_{\rm avg}$ in the most metal-rich galaxies could be due to an increased SFR. This elevated SFR might be itself related to the quick enrichment of the lower-metallicity regions in metal-rich galaxies (see Sect.\,\ref{sec:Zdistrib}).

\subsection{Distribution of physical parameters within galaxies}

The LOC approach is motivated by the study of potential biases due to a single 1D model approach (Sect.\,\ref{sec:locvs1d}) but also because it enables additional parameters relevant to galaxy evolution. While many different model architectures could be used to match the observed lines, we chose a plausible architecture with physical parameters distributed as power laws within each galaxy because they likely represent physically meaningful internal distributions (Sect.\,\ref{sec:ecodecision}). 

\begin{figure*}
\centering
   \includegraphics[width=18cm]{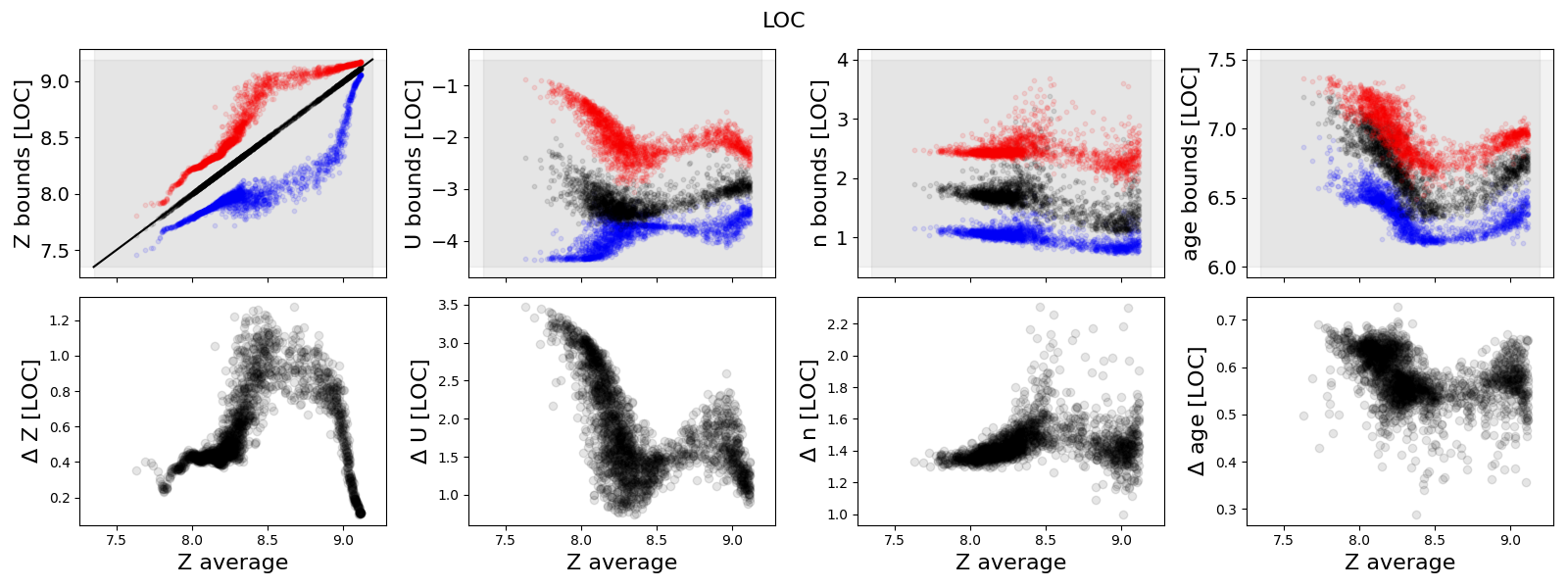}
   \includegraphics[width=18cm]{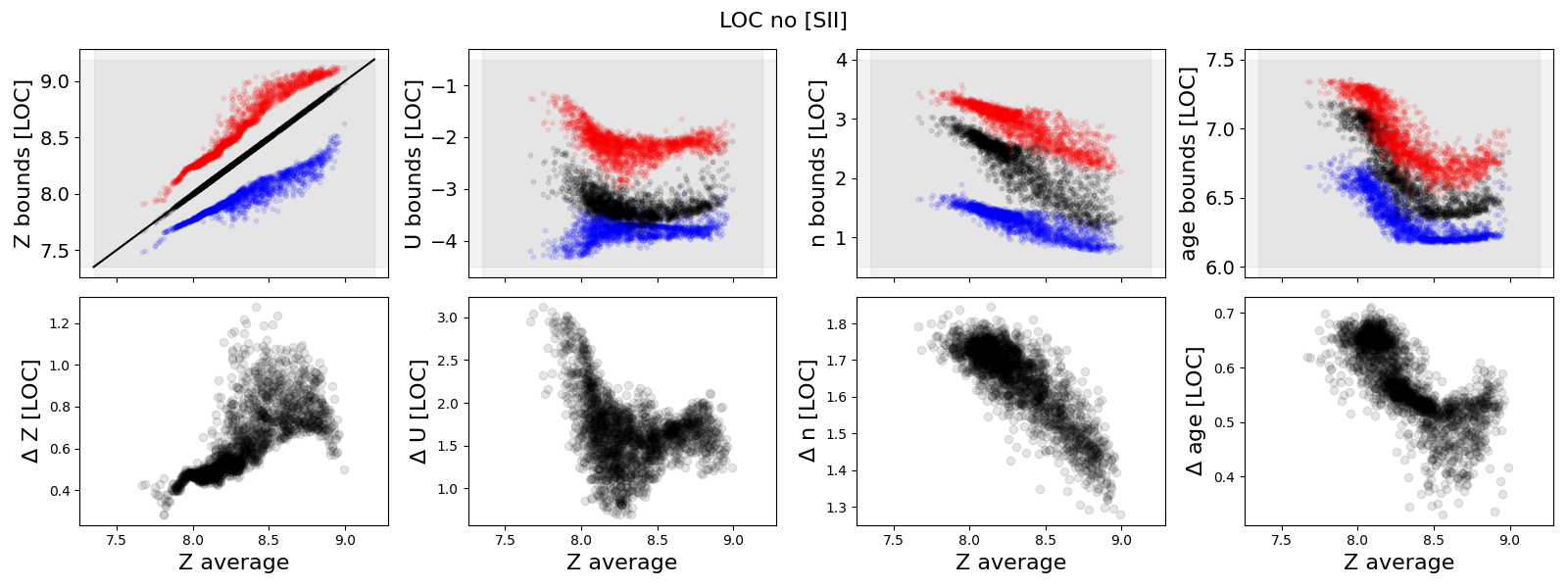}
   \includegraphics[width=18cm]{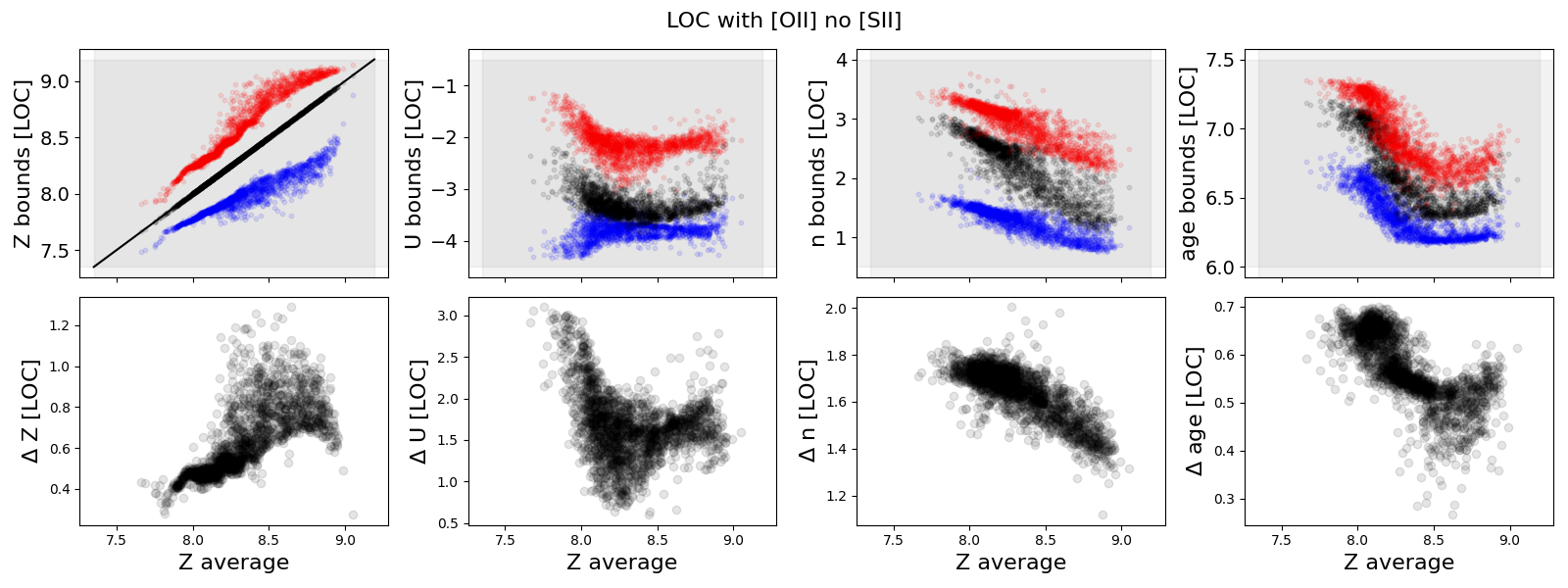}
   \caption{Evolution of the parameter boundaries $p_{\rm min,max}$ versus\ the average metallicity $Z_{\rm avg}$ for the three inference runs (including [S\2] on top, ignoring [S\2] in the middle, and replacing [S\2] by [O\2] in the bottom). For each run, the upper row shows $p_{\rm min,max}$ in blue and red respectively (with the gray rectangles showing the full parameter range in the grid), while the bottom row shows the difference between $p_{\rm max}-p_{\rm min}$.}
   \label{fig:boundaries}
\end{figure*}

In this section we stand by this hypothesis and investigate what these distributions imply as far as galaxy evolution is concerned. In other words, given the observations, given the grid of 1D photoionization models, and given the assumption of power-law distributions, we wish to find and interpret the most likely internal distributions of physical parameters ($Z$, $U$, age, $n$) within each galaxy of the sample. We show in Figure\,\ref{fig:boundaries} the variation of the upper and lower boundaries as well as average values as a function of metallicity discussed in the following. 

\subsubsection{Metallicity internal distribution}\label{sec:Zdistrib}

We first note that the power-law distribution inferred for any given galaxy should not be confused with the metallicity gradient often observed in disk-dominated galaxies (e.g., \citealt{Carton2015a,Hu2018b,Bresolin2019a,Simons2021a}; including in the Milky Way, \citealt{Balser2011a}) and thought to be the result of star-formation spreading outward through the disk (e.g., \citealt{Sharda2021a,Sharda2024a}). A positive or negative slope $\alpha_Z$ in our models does not correspond to the slope of the metallicity gradient (the latter considering radial averages that do not contribute equally to the total emission) but reflects instead a given weight of metal-rich versus\ metal-poor regions emission within a galaxy toward global emission.

The inferred internal metallicity dispersion (${\Delta}Z=Z_{\rm max}-Z_{\rm min}$) reaches up to $\sim0.7-1$\,dex around solar metallicity galaxies (Fig.\,\ref{fig:boundaries}), which is compatible with the dispersion often observed in 2D maps (e.g., \citealt{Poetrodjojo2018a,Nakajima2024a}). However, it must be noted that robust 2D metallicity estimates indicate that the abundance gradient should dominate the metallicity variation (e.g., \citealt{Kreckel2019b,Williams2022a}) and that our result may be the consequence of a 3D distribution and the consequence of potentially larger weights from metal-poor regions contributing to the total luminosity. 

Our results for the ECO sample also show that $Z_{\rm min}$ and $Z_{\rm max}$  do not evolve the same way as a function of $Z_{\rm avg}$  (Fig.\,\ref{fig:boundaries}). We identify 4 regimes based on the metallicity dispersion ${\Delta}Z$, most evident in the inference runs using the [S\2] lines:
\begin{itemize} 
\item[1)] Smooth increase of ${\Delta}Z$ until $\approx1/3$\,Z$_\odot$ by a factor of $\approx2$, with little dispersion across galaxies.
\item[2)] Sharp increase between $\approx1/3$\,Z$_\odot$ and $\approx1/2$\,Z$_\odot$ by a factor of $\approx4$, with a large dispersion across galaxies.
\item[3)] Turnoff until super-solar metallicity galaxies.
\item[4)] Sharp decrease until $\approx2.5-3$\,Z$_\odot$ by a factor of $\approx10$ (only seen if using [S\2] lines as constraints). 
\end{itemize}

We emphasize that the small difference between $Z_{\rm min}$ and $Z_{\rm max}$ in low-metallicity galaxies is a direct result of the inference: other solutions may exist (such as a wide range of metallicity within galaxies together with a very low $\alpha_Z$), but their likelihood is significantly lower. Furthermore, the trend observed for the metallicity dispersion to be small for either the lowest or highest metallicity galaxies is not due to potential edge effects as we do not observe the same behavior for the boundaries of other parameters as a function of the corresponding average parameter value.

In metal-poor galaxies, the small metallicity dispersion implies that the existence of numerous metal-rich regions in low-Z galaxies is unlikely. The relatively slow evolution of $Z_{\rm min}$ may indicate that metal-poor ($\lesssim1/3$\,Z$_\odot$) gas remains present in metal-rich galaxies until at least an average metallicity about solar, but in the form of regions that do not contribute much to the total emission ($\alpha_Z>0$). Inversely, the small dispersion inferred for the most metal-rich galaxies is only seen for inference runs using [S\2] and seems in contradiction with the evidence of relatively metal-poor regions in metal-rich galaxies (e.g., \citealt{Poetrodjojo2018a}). The inference runs ignoring [S\2] do predict a relatively large dispersion instead. 

Assuming that the sample at $z\sim0$ may capture the evolution of galaxies versus\ $Z$ (i.e., assuming a closed-box scenario for which the average metallicity increases monotonously with time and also assuming that metal-poor galaxies are past versions of metal-rich ones), the fact that $Z_{\rm max}$ increases relatively faster (factor of $\approx10$ between $\approx1/3$ and $\approx1/2$\,Z$_\odot$) than $Z_{\rm min}$ might indicate a faster enrichment of metal-rich regions. One possible interpretation, assuming that the average $Z$ traces an evolutionary pathway, is that 
\begin{itemize}
\item[1)] galaxies start forming stars in a gas whose metallicity is relatively uniform and metal-poor (average metallicity below $\lesssim1/5$\,Z$_\odot$), 
\item[2)] star-formation is slightly more efficient in regions already enriched in heavy elements (e.g., due to increased cooling) leading to an increasing offset between the maximum and minimum metallicity within the galaxy and to the average metallicity of the galaxy being driven by metal-rich regions, 
\item[3)] the enrichment of the most metal-rich regions eventually plateaus around solar metallicity, which could be due to the fact that the added metal mass released through a typical star-formation episode becomes small compared to the existing metal content. 
\end{itemize}
A symmetric behavior is observed for $Z_{\rm min}$ and $Z_{\rm max}$, with two reference metallicity thresholds: $1/3$\,Z$_\odot$ corresponding to a sharp increase in the metal enrichment, and $\sim$\,Z$_\odot$ corresponding to a saturation in enrichment.

\subsubsection{Internal distribution of other parameters}

The ionization parameter dispersion within galaxies ($U_{\rm max}-U_{\rm min}$) is the largest within the most metal-poor galaxies and decreases sharply until $\sim1/2$\,Z$_\odot$ (Fig.\,\ref{fig:boundaries}). Since the density boundaries do not evolve much versus\ $Z$, the wide range of $U$ in metal-poor galaxies could be due to a wide range of the ionizing photon flux and/or the distance between the stars and the illuminated gas shells. We remark that the lower boundary drives the average ionization parameter (due to the negative slope $\alpha_U$ (Sect.\,\ref{sec:hyperparameters}).

There is no clear evolution of the density boundaries versus\ $Z$ apart from a slight decrease of both boundaries. As a consequence, there is also little evolution of the density dispersion ($n_{\rm max}-n_{\rm min}$).
The age boundaries tightly follow each other versus\ $Z$ and, as a consequence, the difference depends relatively little on $Z$.

\subsection{On a potential metallicity bimodality}\label{sec:bimodality}

We find a bimodal distribution of the average metallicity $Z_{\rm avg}$ in the galaxies of the ECO sample when using all available lines (Fig.\,\ref{fig:hyperaverages}). The low-metallicity probability peak, where most galaxies lie, is centered around $12+\log (\rm O/H) \approx 8.25$ ($\approx0.3$\,Z$_\odot$) and the high-metallicity peak reaches up to $12+\log (\rm O/H) \approx 9.0$. However, if [S\2] lines are ignored for the inference (for the LOC approach or single 1D model alike), the bimodality is much weaker and the secondary peak lies around the solar value. The bimodality is driven by the upper boundary $Z_{\rm max}$ (Sect.\,\ref{sec:hyperparameters}), while the lower boundary $Z_{\rm min}$ hardly seems to reach the metallicity threshold for rapid enrichment (Sect.\,\ref{sec:Zdistrib}).

The N2S2 empirical diagnostic \citep{Dopita2016a} provides a significantly smoother PDF (Fig.\,\ref{fig:averageZ}) and provides metallicities as low as $\approx1/30$\,Z$_\odot$ while the metallicity we infer does not reach below $1/10$\,Z$_\odot$. We emphasize that the lines involved in the N2S2 diagnostic are not particularly well reproduced by the various models we consider and that the difference seen for the PDFs may be partly due to the inability of the models to reproduce better [S\2] and/or to systematic effects in the line measurement available in the SDSS catalogs (Sect.\,\ref{sec:lineset}). It also shows that the bimodality in $Z$, if real, may be difficult to identify solely based on empirical diagnostics. 

We tested inference runs with different underlying photoionization grids and including the [S\2] lines: BOND \citep{ValeAsari2016a} and SFGX \citep{Ramambason2022a}. Although not shown, the PDFs from SFGX and from the present grid are similar and both show a bimodality. The bimodality is more pronounced with the present grid because the maximum metallicity in SFGX is only $0.1$ log solar. BOND does reach higher values but does not show any bimodality, and in fact provides similar results to the N2S2 calibration. We conclude that the $Z$ bimodality is mostly driven by the grid presently used and the underlying abundance patterns that are assumed (similar prescription in the current grid and SFGX, mostly drawn from \citealt{Nicholls2017a}). Ignoring the [S\2] lines in the inference somewhat mitigates these issues, implying that the bimodality, if real, is likely not a strong one.

For completeness, although the present sample was selected to be compatible with star-forming criteria and negligible AGN contamination (Sect.\,\ref{sec:sample}), we cannot exclude that high-$Z$ galaxies may correspond to sources with a contribution from ionization mechanisms other than UV photoionization from massive stars (e.g., shocks and/or AGN). In fact, the high-$Z$ sample is partly populated with ``ambiguous'' galaxies lying between the SF-AGN demarcation lines of \cite{Kauffmann2003a} and \cite{Stasinska2006a} and may therefore imply weak AGN contamination (see Fig.\,\ref{fig:bpt} and App.\,\ref{sec:sfagn}). The interpretation of a potential metallicity bimodality in star-forming galaxies therefore depends heavily on the maximum AGN contamination allowed, especially if one considers that nuclear activity might never be null. Nevertheless, we do note that the ``ambiguous'' galaxies remain far off the AGN domain in the [O\1] diagnostic plot (Fig.\,\ref{fig:bpt}).

\subsection{Mass-metallicity relationship}

The ECO star-forming galaxy sample provides an opportunity to study the mass-metallicity relationship (MZR; see review in \citealt{Maiolino2019a}) in a consistent way for a wide range of galaxy mass. We use the metallicity inferred in the present study and the stellar mass is taken from \cite{Hutchens2023a}.

\begin{figure}
\centering
   \includegraphics[width=9cm]{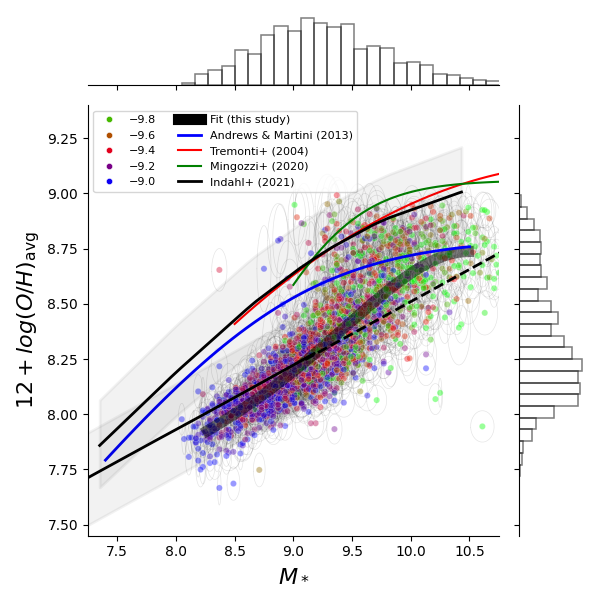}
   \caption{Metallicity-mass relationship using the metallicity  inferred ignoring the [S\2] lines. The thick gray curve shows the 4th order polynomial fit (see text). The blue, red, and green curves show the correlations from \cite{Andrews2013a}, \cite{Tremonti2004a}, and \cite{Mingozzi2020a} respectively, while the bottom and top stripes show the low-mass galaxy fit and the SDSS star-forming galaxy fit from \cite{Indahl2021a} respectively. The dashed line shows the extrapolation from  \cite{Indahl2021a}. }
   \label{fig:correlations_mstar}
\end{figure}

Figure\,\ref{fig:correlations_mstar} shows that the MZR is smooth and narrow until stellar masses $10^9$\,M$_\odot$, but there is a large spread of metallicities in the range $10^{9.5-10}$\,M$_\odot$, which eventually leads to a high-metallicity plateau for the most massive galaxies. The $Z$ and $M_*$ PDFs are remarkably different, the former being somewhat bimodal (Sect.\,\ref{sec:bimodality}) and the latter showing a single-peak broad distribution.

\begin{figure*}
\centering
   \includegraphics[width=9cm]{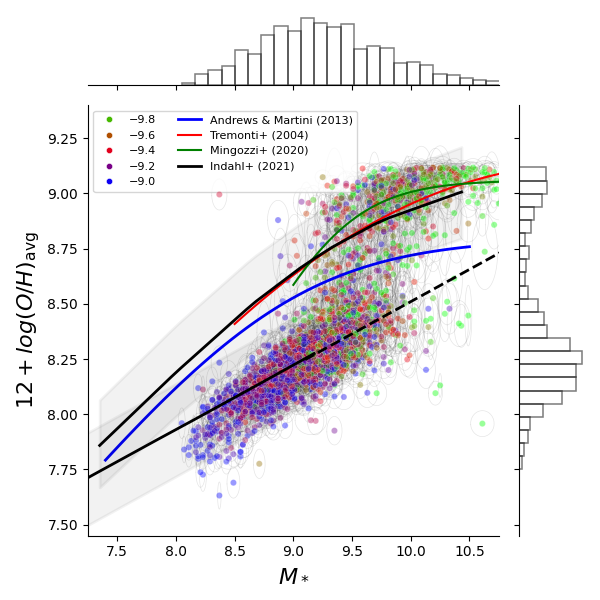}
   \includegraphics[width=9cm]{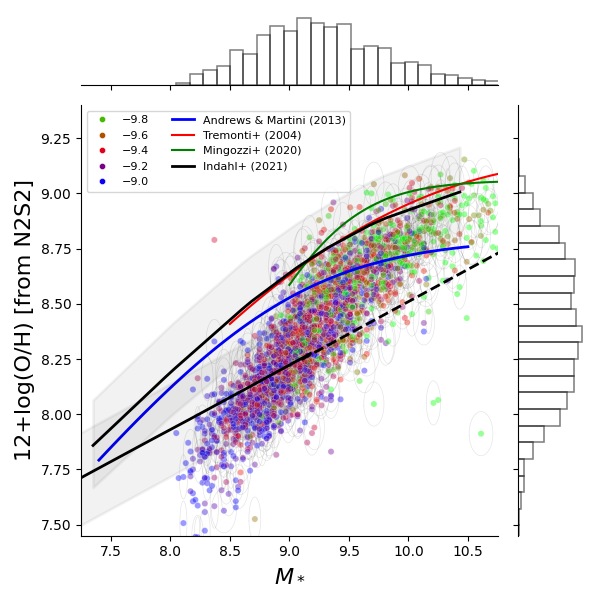}
   \caption{Metallicity-mass relationship using the metallicity inferred with MULTIGRIS including [S\2] lines (left) and using the metallicity calculated from N2S2 empirical calibration (right). }
   \label{fig:correlations_mstar_xtras}
\end{figure*}

The main result shows that the inferred MZR in the low-mass regime ($\lesssim10^{9.5}$\,M$_\odot$, i.e., the bulk of the ECO sample) is compatible with the low-mass galaxy fit in \cite{Indahl2021a}, which uses the robust direct method ($T_e$-method with calibrated ionization correction factors; see also \citealt{Berg2012a,Kirby2013a}). This agreement with \cite{Indahl2021a} as well as with stellar abundances in \cite{Maiolino2019a} suggests that the inferred metallicity (i.e., using photoionization models and strong lines) are reliable.
Figure\,\ref{fig:correlations_mstar_xtras} (left panel) shows that the low-mass fit remains unchanged whether [S\2] lines are used as constraints or not. Figure\,\ref{fig:correlations_mstar_xtras} (right panel) shows that the MZR in the low-mass regime using the empirical N2S2 diagnostic agrees less well with \cite{Indahl2021a}, suggesting that the [S\2] line measurement may lead to systematics (Sect.\ref{sec:lineset}).

Concerning high-mass galaxies ($\gtrsim10^{9.5}$\,M$_\odot$), we find significant lower metallicities than both \cite{Tremonti2004a} and \cite{Mingozzi2020a}, the latter studies using strong lines and the theoretical method (stellar population + photoionization grids). We argue that this may be a consequence of ignoring the [S\2] lines for inference in the present study, as including these lines results in significantly higher metallicities (Fig.\,\ref{fig:correlations_mstar_xtras} left panel), at the expense of a strong metallicity bimodality. Our fit for high-mass galaxies ignoring [S\2] lines is in line with the study of \cite{Andrews2013a} which uses the direct method on stacks, strengthening the reliability of our inferred metallicity across the full mass range.

In summary, the MZR we infer without [S\2] lines is in line with studies using the direct method from the calibrated range up to methods using stacks, hinting that our models are able to capture the physical conditions of the gas. Our sample was indeed selected to ensure sufficient S/N in the strongest lines but we did not consider $T_e$-sensitive auroral lines (Sect.\,\ref{sec:sample}). Among these, the [O\3] $\lambda4363$ line is detected in only $\approx17\%$ and $\approx7\%$ of the sample above $2\sigma$ and $3\sigma$ respectively. Nevertheless, we have verified that the model predictions for this line (i.e., not using it for inference) agree within $2\sigma$ for the galaxies with detections. We also verified that using it for the inference does not modify our results across the mass and metallicity ranges.

The fourth order polynomial fit, valid in the range $\log M*=[8.25, 10.5]$, provides
\begin{equation}
  \begin{aligned}
    12 + \log({\rm O/H}) \approx -0.035756 M_*^4  + 1.25737 M_*^3 - 16.4913 M_*^2 \\
    + 95.9941 M_* - 201.99.
    \end{aligned}
  \end{equation}

  The MZR is often reported in the literature to show a transition between a positive correlation to an almost constant metallicity, thought to be the consequence of galactic outflow efficiency versus\ galaxy mass (e.g., \citealt{Tremonti2004a,DeVis2019a}), i.e., with no gap or sharp transition between two metallicities. A bimodality in galaxy parameters is, however, known to exist between blue star-forming disks and red spheroids dominated by old stellar populations, with a mass transition $\approx10^{10.5}$\,M$_\odot$ and has been attributed to cold flows and shock heated streams (e.g., \citealt{Dekel2008a}).
  
Various studies have proposed ``inverse'' morphological transformations (from early- to late-type) through a disk regrowth process possibly enabled by gas accretion, which may explain the existence of non-cluster “blue-sequence” E/S0 galaxy population as well as extended UV emission around some early-type galaxies (see, e.g., \citealt{Stark2013a} and \citealt{Moffett2015a}). This led \cite{Kannappan2013a} to hypothesize that 
the transition is due to a different refueling regime with high levels of external gas accretion and stellar mass growth. The “blue-sequence” E/S0 galaxy population a population exists primarily below a stellar mass of $\sim10^{9.7-10.5}$\,M$_\odot$ and corresponds well to the masses for which the average metallicity we infer increases sharply, potentially suggesting a higher star-formation efficiency. The present results unfortunately do not allow us to distinguish between a higher star-formation efficiency due to external processes (accretion) or internal processes (metallicity threshold; Sect.\,\ref{sec:Zdistrib}).


\section{Conclusions}

We present models of star-forming galaxies from the volume-limited ECO catalog. The main objective is to interpret realistic models of an unbiased sample in order to probe relationships between physical parameters, in particular as a function of metallicity, but also to investigate the metallicity probability density function itself and to recover the internal distribution of physical parameters within galaxies. In summary:
\begin{itemize}
\item We designed a framework using probabilistic methods in order to assess various model architectures meant to describe the emitting components of a galaxy. In particular, we considered the combination of many 1D models, that is, the LOC hypothesis. LOC architectures integrate a number of models with different physical properties linked by a given distribution (e.g., a power law) whose parameters are found through inference.
\item We applied this framework to the ECO star-forming galaxy sample. We focused on a few model architectures, including a single 1D model approach for comparison. The 1D models, used as single models or within an LOC combination, were computed with Cloudy with specific abundance patterns as a function of metallicity and we made use of the stellar population synthesis code BPASS.
\item Guided by potential issues with the line measurements as well as with the model hypotheses, we performed runs ignoring [S\2], which globally performed much better and alleviated some issues with [N\2] and [O\1] predictions. 
\end{itemize}

The main results are as follows:
\begin{enumerate}
\item Globally, we find that the LOC models outperform the single 1D models, strengthening the need for relatively complex and realistic architectures. The single 1D models provide values for physical parameters that are close to the average value considering a distribution of components (LOC) -- which we consider robust -- despite the small biases observed and discussed. 
\item For LOC models, the average physical parameter value in a galaxy is always tightly constrained. Other distribution hyperparameters (the slope and boundaries) are much less well constrained but do show small deviations from galaxy to galaxy and with respect to the prior, suggesting that it is possible and meaningful to study these variations for a physical interpretation.
\item We find, in particular, that the integrated emission of galaxies is dominated by relatively low-excitation gas, with an average $U\sim-3.2$. The age distribution peaks around $5$\,Myr, with the lower and upper boundary around $2$ and $10-20$\,Myr, respectively. 
\item The average metallicity shows a weakly bimodal distribution, with most galaxies showing an average metallicity of $\sim0.3$\,Z$_\odot$ and a secondary peak around solar metallicity. 
\item The lower and upper metallicity boundaries within galaxies do not evolve the same way as a function of the average metallicity. In the most metal-poor galaxies, most emitting components have the same metallicity within a factor of $2-3$. As the metallicity increases until about solar values, the most metal-rich regions increase their metallicity sharply while low-metallicity regions remain constant, resulting in a metallicity dispersion up to a factor $5-10$. For super-solar metallicity galaxies, the most metal-poor regions finally get enriched. We propose that this reflects an evolutionary sequence involving a combination of metallicity thresholds for efficient star formation ($\approx1/3$\,Z$_\odot$) and saturation ($\approx$\,Z$_\odot$).
\item The average metallicity bimodality is driven by the upper boundary $Z_{\rm max}$ and the secondary peak could be a consequence of efficient or rapid enrichment of the most metal-rich regions. 
\item We find correlations between all parameters (age, ionization parameter, and -- though to a lesser extent -- density) versus\ $Z$, with the lowest metallicity galaxies having a younger age, higher density, and higher ionization parameter. We find, however, a flattening of age and $U$ for galaxies above $\sim0.5$\,Z$_\odot$. 
\item Finally, we examined the MZR and find results in line with direct abundance method determinations, from the calibrated range at a low metallicity to methods using stacks at a high metallicity. This suggests that the models are able to capture physical conditions of the gas and that the inferred metallicity is reliable. We identified two regimes, the low-mass regime below $\sim10^{9.5}$\,M$_\odot$, reproducing the low-mass galaxy fit from \cite{Indahl2021a}, and a sharp metallicity increase for more massive galaxies. This transition may be related to a specific refueling of non-cluster early-type galaxies but we cannot exclude purely internal processes such as a metallicity threshold for efficient star formation. 
   \end{enumerate}
   
\begin{acknowledgements}
This work is supported by the FACE Foundation Transatlantic Research Partnership Fund (award TJF21\_053). CR gratefully acknowledges the support of the Elon University Japheth E. Rawls Professorship, Elon University's FR\&D committee, and the Advanced Cyberinfrastructure Coordination Ecosystem: Services \& Support (ACCESS), which is supported by the National Science Foundation. This work used the ACCESS resource Expanse at the San Diego Supercomputing Center through allocation AST140040. LR gratefully acknowledges funding from the Deutsche Forschungsgemeinschaft (DFG, German Research Foundation) through an Emmy Noether Research Group (grant number CH2137/1-1).
\end{acknowledgements}

\bibliography{bibtexendum} 

\begin{appendix}

\onecolumn

\section{Empirical metallicity diagnostics and star-formation / AGN demarcation line}\label{sec:sfagn}

\begin{figure}
\centering
   \includegraphics[width=9cm]{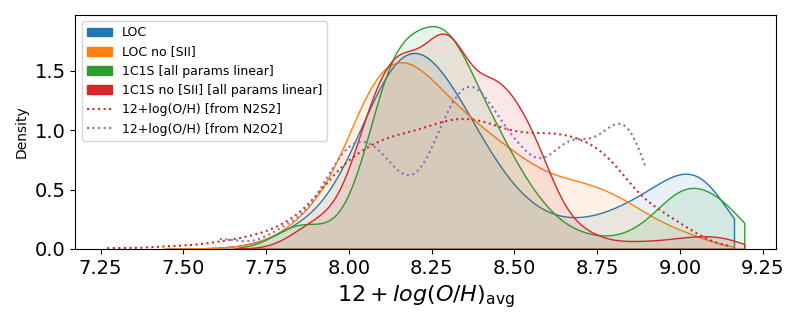}
   \includegraphics[width=9cm]{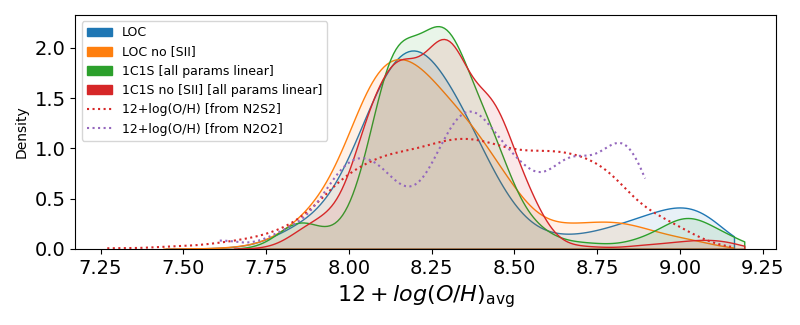}
   \caption{Left: average metallicity PDF including the N2S2 and N2O2 calibrations from \cite{Garg2024a}. Right: same as left plot but using the star-formation / AGN demarcation line of \cite{Stasinska2006a} instead of \cite{Kauffmann2003a}.}
   \label{fig:averageZ}
\end{figure}

\begin{figure}
\centering
   \includegraphics[width=9cm]{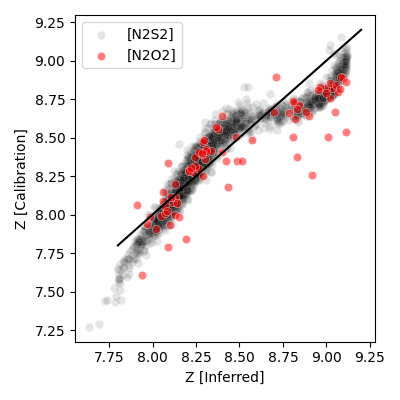}
   \includegraphics[width=9cm]{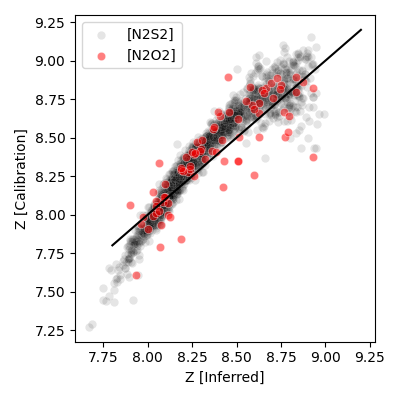}
   \caption{N2S2 (black) and N2O2 (red) empirical diagnostics versus\ the inferred (average) metallicity for the LOC architecture using [S\2] lines (left) or ignoring them (right). }
   \label{fig:ZDopita}
 \end{figure}

Figure\,\ref{fig:averageZ} shows the metallicity distribution inferred through various model architectures compared to the metallicity calculated from empirical diagnostics (N2S2 and N2O2). The N2S2 diagnostic is discussed in Sect.\,\ref{sec:bimodality}. The N2O2 diagnostic suffers from the weak statistics due to the low S/N in the [O\2] line for most of the sample. Figure\,\ref{fig:ZDopita} illustrates how the metallicity we infer differs from the N2S2 and N2O2 empirical diagnostics. 

The sample used in this study was drawn from the ECO catalog, selecting only star-forming galaxies (Sect.\,\ref{sec:sample}). As explained in the main text, the exact choice of the demarcation line between gas excitation dominated by star-formation or not has some impact on the results. \cite{Stasinska2006a} provided an updated demarcation compared to \cite{Kauffmann2003a} to account for galaxies with weak AGN contribution (typically $\lesssim3$\%). We have verified that the results presented in this study remain unchanged whatever the choice of the demarcation (in other words including or not galaxies with potentially weak AGN contributions). Most galaxies falling between the two demarcation lines are high-metallicity galaxies (Fig.\,\ref{fig:bpt}) and the main impact of using the demarcation from \cite{Stasinska2006a} instead of \cite{Kauffmann2003a} is to reduce the statistics of the high-$Z$ galaxies in the various plots, with no change to the actual trends. Unsurprisingly, the only diagnostic plot which changes significantly is the metallicity PDF itself. Results in Section\,\ref{sec:bimodality} use the demarcation from \cite{Kauffmann2003a} and indicate a potential secondary high-$Z$ ``peak'' (highly populated if using [S\2] lines for inference, weakly populated if ignoring [S\2]). When using \cite{Stasinska2006a} instead, we see that the secondary peak is much smaller (Fig.\,\ref{fig:averageZ}).  

\section{Results for individual galaxies}\label{sec:singleresults}

\begin{figure}
\centering
   \includegraphics[width=9cm]{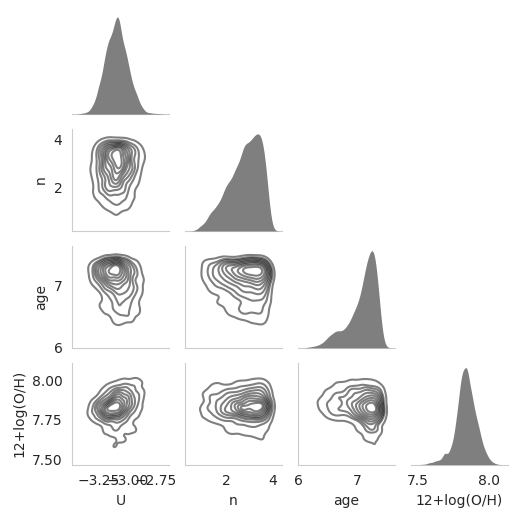}
   \caption{Illustration of the pairwise relationships between the inferred average parameter values for a low-metallicity source, illustrating that the posterior distributions do not show strong correlations.}
   \label{fig:illustration_single_pairplot}
 \end{figure}

Figure\,\ref{fig:illustration_single_pairplot} shows that the inferred average parameter values for each individual galaxy show no significant correlations or degeneracy within the confidence intervals. The inference method in MULTIGRIS makes use of the Sequential Monte-Carlo method from python package PyMC \citep{Salvatier2016a}, which is well adapted to complex, potentially multimodal posterior distributions (see \citealt{Lebouteiller2022a}). We have used the largest possible number of particles to alleviate issues with minor modes being cropped. 

\begin{figure*}
\centering
   \includegraphics[width=19cm,height=7cm]{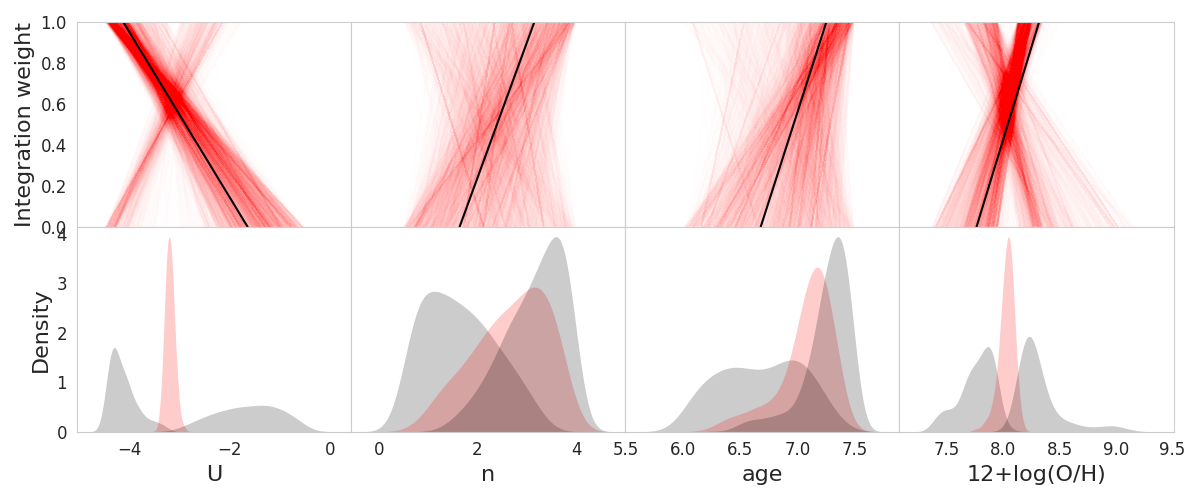}
   \includegraphics[width=19cm,height=7cm]{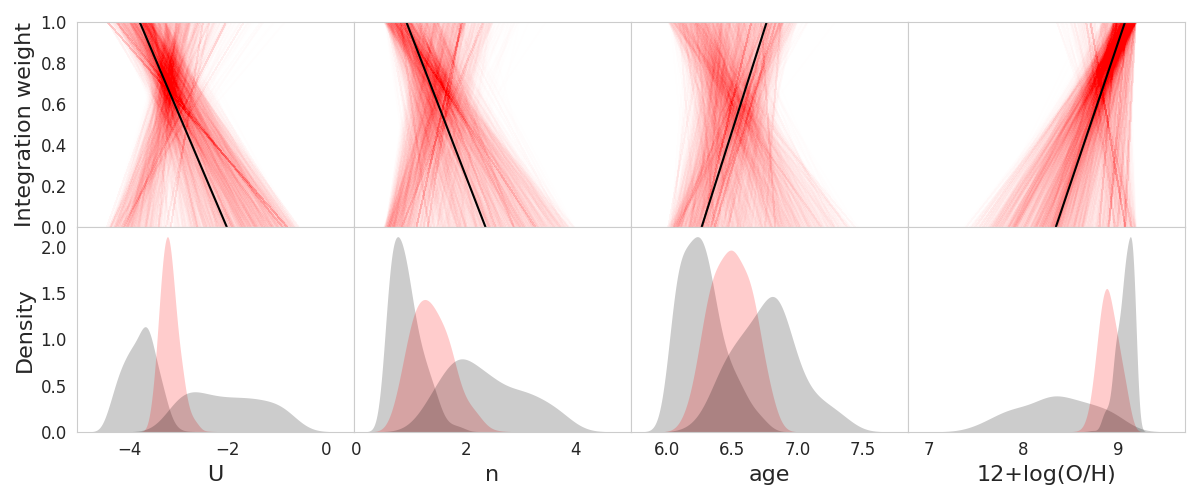}
   \caption{Illustration of inferred power-law distributions for typical subsolar metallicity (on top) and super-solar metallicity (on bottom) galaxies. For each metallicity case we show the individual draws for the integration weight ($\Phi(p)$; Eqn.\,\ref{eq:alpha}) on top and the PDF for the lower and upper boundaries for integration ($p_{\rm min,max}$; gray) and the average parameter value ($p_{\rm avg}$; red) the bottom).}
   \label{fig:illustration_single}
 \end{figure*}

Examples of inference results for individual galaxies are shown in Figure\,\ref{fig:illustration_single}. The average $U$ and $Z$ values within any galaxy (defined in Eqn.\,\ref{eq:avg}) are generally well constrained, and some boundaries (e.g., upper boundary $U_{\rm max}$) are relatively uncertain. However, it is worth noting that there is often minimal overlap between the lower and upper boundaries for each parameter, reflecting the fact that, given the choice between a single valued parameter or an LOC distribution, the latter is always preferred and thus likely to represent a more realistic model architecture.

\begin{figure*}
\centering
   \includegraphics[width=9cm]{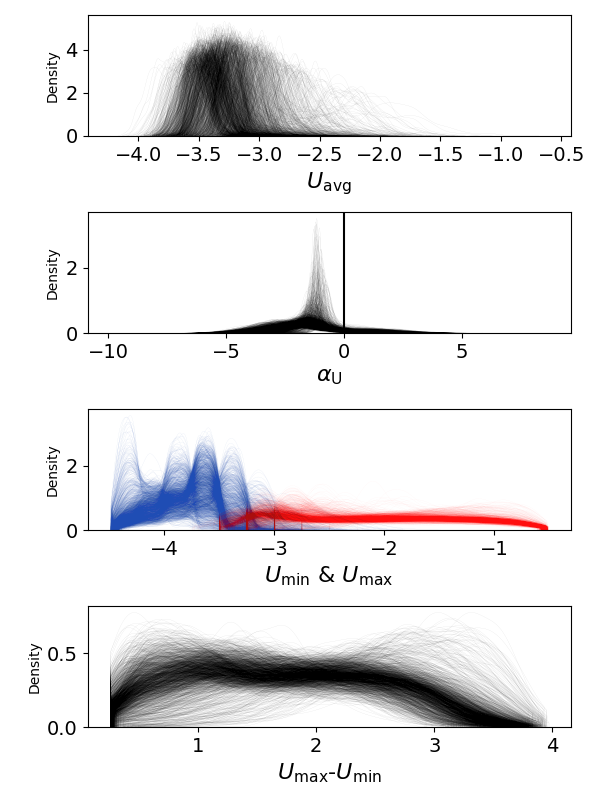}
   \includegraphics[width=9cm]{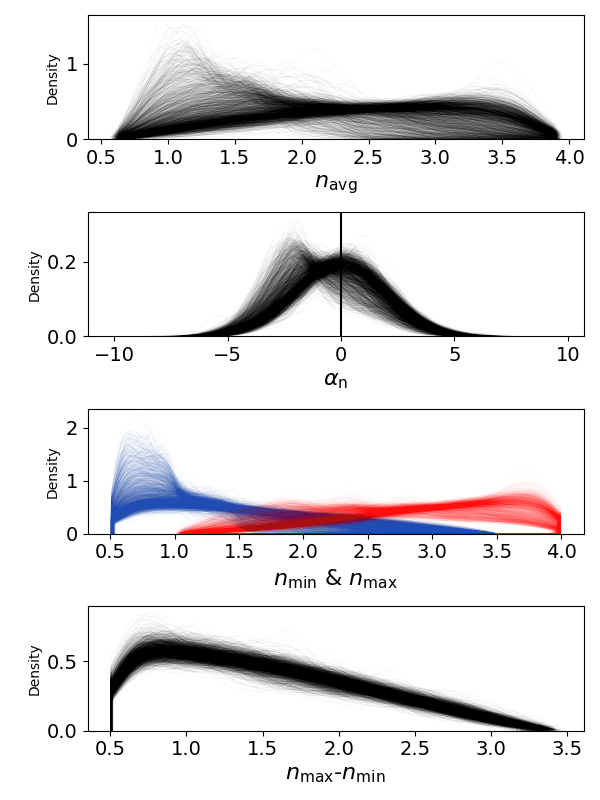}   \includegraphics[width=9cm]{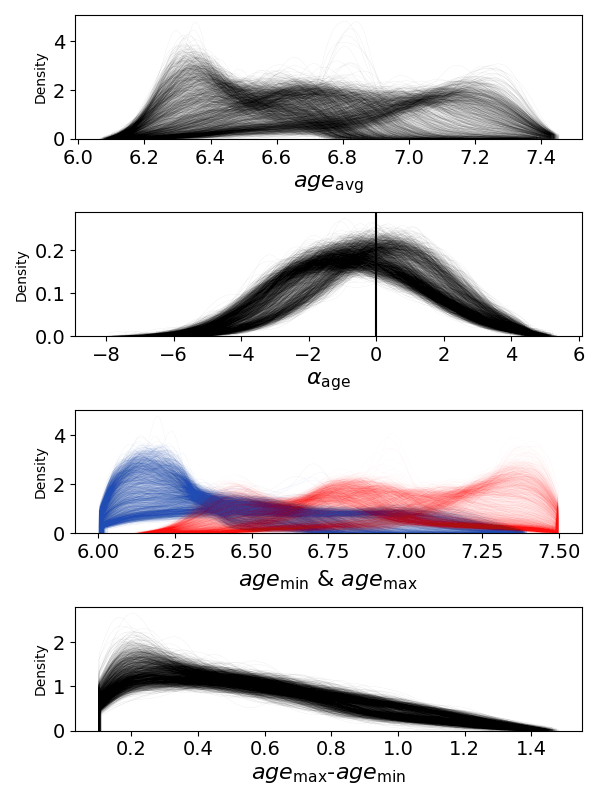}
   \includegraphics[width=9cm]{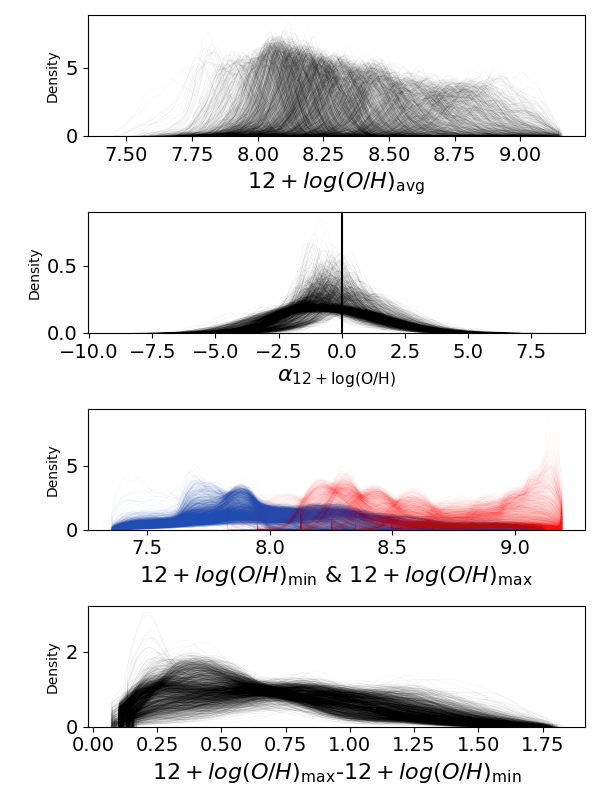}
   \caption{Illustration of the parameter distribution for each individual galaxy (here LOC approach ignoring [S\2] lines; see Sect.\,\ref{sec:lineset}). Each individual curve represents the PDF of a single galaxy. We plot  the lower and upper boundaries  in different colors for clarity. }
   \label{fig:params_indiv}
\end{figure*}

Figure\,\ref{fig:params_indiv} shows the parameter distribution for each galaxy (not to be confused with the parameter distribution of the average values, shown in Fig.\,\ref{fig:hyperaverages}). This figure shows that the PDF of some parameters is driven by the actual variation between galaxies (e.g., average metallicity) while the PDF of some other parameters is somewhat driven by a similar PDF inferred for all galaxies (e.g., age dispersion).

\end{appendix}

\end{document}